# Ultrafast dynamics of CN radical reactions with chloroform solvent under vibrational strong coupling

Ashley P. Fidler,[1] Liying Chen,[1] Alexander M. McKillop,[1] and Marissa L. Weichman[1,a)]

[1]Department of Chemistry, Princeton University, Princeton, New Jersey 08544, USA

a)weichman@princeton.edu

**ABSTRACT**

Polariton chemistry may provide a new means to control molecular reactivity, permitting remote, reversible modification of reaction energetics, kinetics, and product yields. A considerable body of experimental and theoretical work has already demonstrated that strong coupling between a molecular vibrational mode and the confined electromagnetic field of an optical cavity can alter chemical reactivity without external illumination. However, the mechanisms underlying cavity-altered chemistry remain unclear in large part because the experimental systems examined previously are too complex for detailed analysis of their reaction dynamics. Here, we experimentally investigate photolysis-induced reactions of cyanide (CN) radicals with strongly-coupled chloroform ($CHCl_3$) solvent molecules and examine the intracavity rates of photofragment recombination, solvent complexation, and hydrogen abstraction. We use a microfluidic optical cavity fitted with dichroic mirrors to facilitate vibrational strong coupling (VSC) of the C−H stretching mode of $CHCl_3$ while simultaneously permitting optical access at visible wavelengths. Ultrafast transient absorption experiments performed with cavities tuned on- and off-resonance reveal that VSC of the $CHCl_3$ C−H stretching transition does not significantly modify any measured rate constants, including those associated with the hydrogen abstraction reaction. This work represents, to the best of our knowledge, the first experimental study of an elementary bimolecular reaction under VSC. We discuss how the conspicuous absence of cavity-altered effects in this system may provide insights into the mechanisms of modified ground state reactivity under VSC and help bridge the divide between experimental results and theoretical predictions in vibrational polariton chemistry.

## I. INTRODUCTION

Photonic control of molecular processes is a long-standing goal of physical chemistry with the potential to revolutionize chemical synthesis.[1, 2] Photonic control schemes have historically used targeted electromagnetic fields to steer chemical reactions through a preferred pathway, enabling selective chemical transformations without the need for synthetic modifications to the molecular structure or environment.[2, 3] One such approach, vibrational mode-selective chemistry, uses narrowband laser excitation to pump specific molecular vibrations relevant to the reaction coordinate and thereby steer the system towards a particular product channel.[4-7] In parallel, coherent control schemes using femtosecond pulse shaping techniques can launch carefully designed wavepackets to influence chemical reaction trajectories.[8-10] However, broader applications of both mode-selective chemistry and coherent control are spoiled by the rapid dissipation of localized vibrational energy to anharmonically coupled degrees of freedom via intramolecular vibrational energy redistribution (IVR) and to the surrounding environment via intermolecular energy transfer (IET).[2, 3, 11] As a result, only simple, predominantly gas-phase systems have proven amenable to optically-driven chemical control schemes.

The emerging field of polariton chemistry may soon provide the robust and flexible architecture necessary for photonic control of more complex, solution-phase systems.[12-19] In contrast to laser-based control techniques, polaritonic schemes harness strong light-matter interactions engineered in optical cavities to alter chemical reactivity. Here, we investigate the photolysis-initiated bimolecular reaction dynamics of molecules whose vibrational modes are strongly coupled to an infrared (IR) optical cavity. Inspired by reports of modified hydrogen abstraction rates following selective excitation of C−H bonds in the gas-phase mode-selective chemistry literature,[20, 21] we examine prospects for using vibrational strong coupling (VSC) to alter the reactivity of C−H bonds in solution. Here, we target the elementary CN + CHCl$_3$ hydrogen abstraction reaction system in particular. This theoretically-tractable reaction serves as testbed to interrogate open questions in cavity-altered reaction mechanisms in a regime not previously realized experimentally.

Polariton formation is a consequence of strong light-matter coupling between a bright optical transition of a molecular ensemble and the confined electromagnetic field of an optical cavity (Fig. 1(a)).[12-19] The collective strong coupling regime is reached when the coherent energy exchange rate between the molecular and photonic modes exceeds the dephasing rate of each individual constituent.[14] As described by cavity quantum electrodynamics, collective strong coupling changes the eigenstates of the coupled system, generating two new hybrid light-matter states called polaritons separated in energy by the Rabi splitting ($\hbar\Omega_R$), as well as a manifold of dark states lying at the energy of the original molecular resonance (Fig. 1(b)).[22, 23] According to the Tavis-Cummings model, the magnitude of the collective Rabi splitting scales as $\sqrt{N/V}$ in a cavity of volume $V$ containing $N$ coupled absorbers.[23-25] One can therefore more readily access the strong coupling regime in dense samples, making solution-phase systems ideal targets for polaritonics. Moreover, cavity-mediated strong light-matter coupling can occur in the dark, as enclosed molecules couple with the cavity vacuum field.[26] This unusual phenomenon may enable constant baseline modification of molecular behavior simply by enclosure in a resonant cavity, unlike conventional laser-based control experiments that rely on precisely-timed, intense fields to change chemical reaction outcomes.

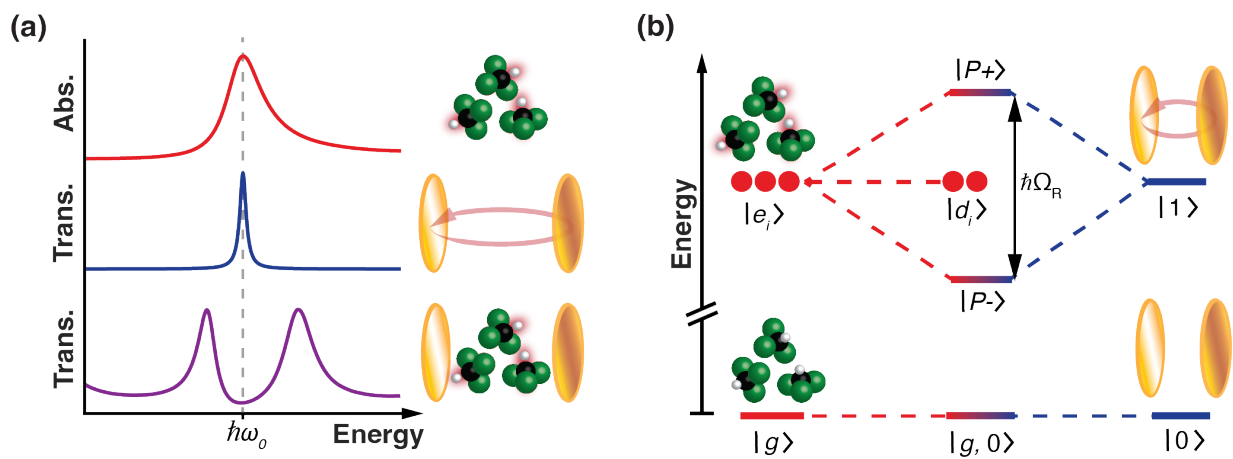

**FIG. 1.** Strong light-matter coupling of a molecular ensemble in a Fabry-Pérot optical cavity. **(a)** Strong coupling manifests in cavity transmission spectra as two new features (purple) split by a frequency greater than the linewidths of both the molecular absorption feature (red) and the cavity mode (blue) resonant at $\hbar\omega_0$. **(b)** Schematic representation of collective strong coupling as described by the Tavis-Cummings model. When molecules confined in an optical cavity possess a sufficiently bright molecular transition ($|g\rangle \rightarrow |e_i\rangle$) with a transition frequency degenerate with excitation of a quantized cavity mode ($|0\rangle \rightarrow |1\rangle$), upper ($|P+\rangle$) and lower ($|P-\rangle$) polariton states form separated by the Rabi splitting ($\hbar\Omega_R$), alongside a manifold of uncoupled dark states ($|d_i\rangle$).

While polaritons have been studied for decades as a consequence of cavity quantum electrodynamics in the semiconductor[27] and atomic physics communities,[28, 29] strong coupling has only recently been explored in molecular systems relevant to chemistry.[12-15, 18, 26] Studies examining strong coupling of electronic resonances have seen evidence of cavity-modified photophysical properties[30-32] and photochemical reaction rates,[33-35] although questions remain about the origin of these effects.[36] Similarly, strong coupling of molecular vibrational modes has been widely studied in pursuit of modifying chemistry in the electronic ground state.[15, 18, 19, 37, 38] Following Ebbesen and coworkers' early reports of significant rate reductions[39] and altered branching ratios[40] in silane deprotection reactions under VSC, modified ground state reactivity has been reported in a number of other systems, including Prins cyclization,[41] Woodward-Hoffman ring opening,[42] phenyl isocyanate alcoholysis,[43] and ATP hydrolysis.[44] Despite lingering concerns about the reproducibility of certain results reporting cavity catalysis,[45, 46] these experimental demonstrations have galvanized considerable theoretical effort to identify the mechanisms by which VSC modulates chemical reactions.[12, 17, 23, 38, 47] While theorists have proposed explanations ranging from altered dark state dynamics[48, 49] to cavity-mediated energy redistribution[50-54] to dynamical caging by cavity photons,[55, 56] the exact mechanisms underlying vibrational cavity-altered chemistry remain unclear. This lack of resolution can be attributed in part to the discrepancy between the available theoretical tools and the complex, solution-phase systems currently documented in the experimental VSC literature. The vast majority of reactions reported under VSC thus far proceed thermally over substantial activation barriers on timescales of minutes to hours, and have been probed primarily by indirect measurements without quantum-state-specific resolution.[18] New laboratory work must strive to meet theory halfway, targeting simpler systems to validate and benchmark predictive models for cavity-altered chemistry.

Here, we aim to help bridge this experimental-theoretical divide by extending polariton chemistry into a new category of theoretically-tractable solution-phase chemical reactions: hydrogen abstraction reactions involving strongly-coupled solvent molecules. In this initial demonstration, we utilize electronic ultrafast transient absorption (TA) spectroscopy to examine the CN + CHCl$_3$ → HCN + CCl$_3$ hydrogen abstraction reaction under VSC of the chloroform C−H symmetric stretching mode (Fig 2(a)). This benchmark system offers distinct advantages as a probe of polariton chemistry. As a highly exothermic elementary reaction ($\Delta H^0 = -120$ kJ/mol)[57] that is essentially barrierless in solution, the CN + CHCl$_3$ hydrogen abstraction reaction scheme has been extensively characterized outside of cavities both theoretically[58-60] and experimentally.[61-65] We can rely on convenient, pre-existing spectroscopic schemes to directly track the reaction dynamics of CN radicals via their electronic transient absorption signatures,[61-64] providing a solid foundation from which to screen for cavity-altered behavior. In addition, the optically bright C−H stretching mode of CHCl$_3$ is an attractive candidate for VSC. CHCl$_3$ is present at high concentration as both solvent and co-reactant, enabling a sufficient collective coupling strength to experimentally access the VSC regime. We employ an optical cavity with dichroic mirrors that facilitate VSC at infrared wavelengths while maintaining optical access at ultraviolet and visible (UV-Vis) wavelengths, empowering us to take advantage of direct, electronic state-specific measures of reaction progress in order to determine whether these dynamics are altered by VSC.

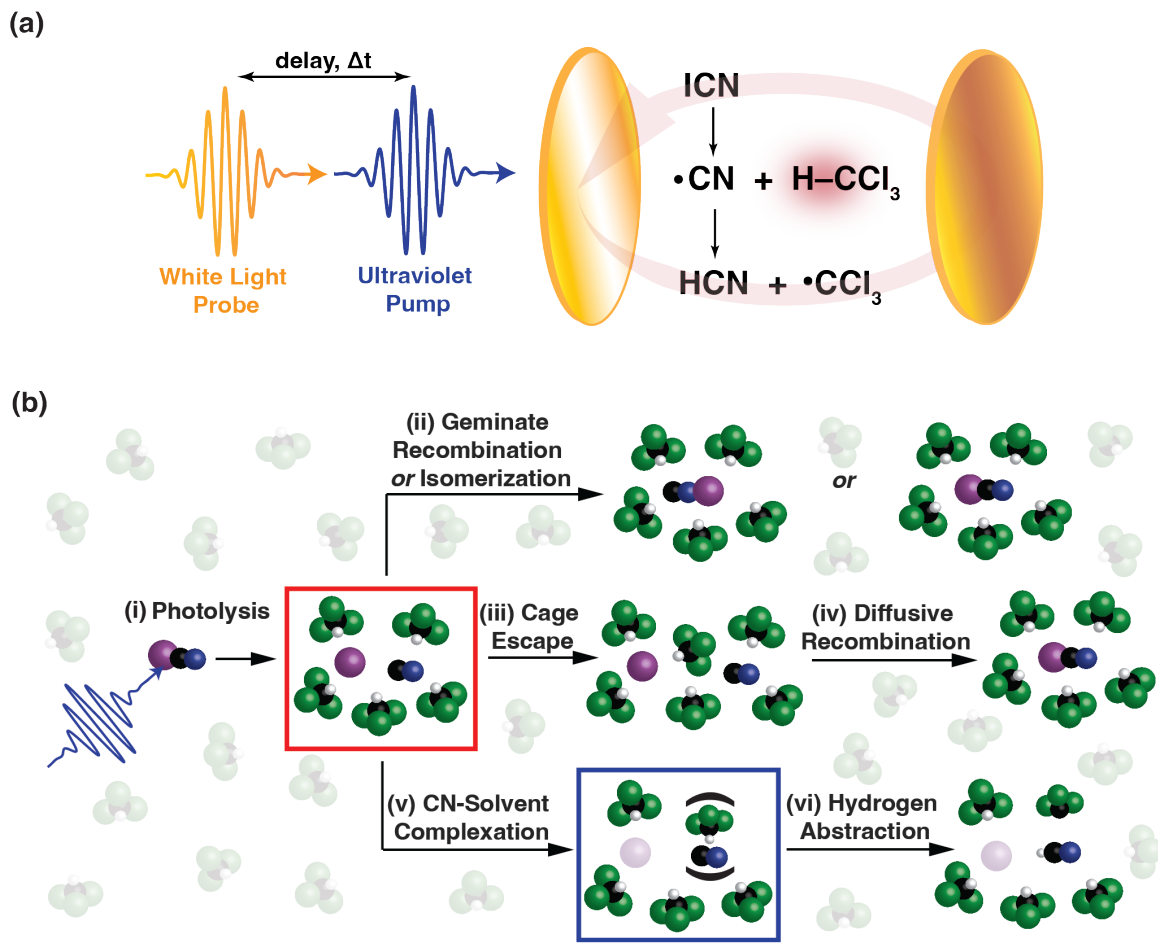

**FIG. 2.** CN radicals undergo various reactive processes in CHCl$_3$ solvent following photolysis of ICN. **(a)** We probe intracavity hydrogen abstraction reactions of CN radicals with strongly-coupled chloroform solvent using ultraviolet pump – white light probe transient absorption spectroscopy. **(b)** Schematic depicting the fate of CN radicals in more detail. Following photolysis of the ICN precursor (i), the initial free CN radical population can decay via multiple pathways as shown in (ii) – (vi). Transient absorption experiments monitor the decay of the initial free CN radical signature at 390 nm (red box) and the formation and loss of the CN-solvent complex signature at 340 nm (blue box).

Following prior work from the solution-phase molecular dynamics literature,[64] we initiate reactions by producing CN radicals *in situ* via ultrafast photolysis. A schematic depicting the subsequent fates of these species is presented in Fig. 2(b). We use an ultraviolet (UV) pump pulse to excite ground state $\tilde{X}\ ^1\Sigma^+$ ICN molecules to repulsive states that comprise the $\tilde{A}$ band absorption continuum, initiating dissociation into ground state $\tilde{X}\ ^2\Sigma^+$ CN radicals and atomic iodine radicals in both the ground I($^2P_{3/2}$) and spin-orbit excited I*($^2P_{1/2}$) states.[58, 66, 67] The free CN radical population gives rise to a relatively sharp spectral feature centered near 390 nm, corresponding to the CN $\tilde{B} \leftarrow \tilde{X}$ electronic transition.[62, 68] Strong interactions with solvent are known to shift this CN radical absorption feature to higher frequencies.[66, 68] In CHCl$_3$, the 390 nm feature exhibits only a minor blue-shift from the 388 nm gas-phase CN $\tilde{B} \leftarrow \tilde{X}$ absorption maximum, suggesting that the nascent CN radicals interact only weakly with the surrounding CHCl$_3$ solvent cage.[61, 67-69] More strongly-bound CN-solvent complexes[63] form over time, as evidenced by a broad feature centered near 340 nm that arises at later pump-probe delays. By monitoring TA signatures at both 390 nm and 340 nm following photolysis, we can simultaneously recover rates of geminate and diffusive photofragment recombination, solvent complexation, and hydrogen abstraction under a variety of extracavity and intracavity coupling conditions.

Previous work in the polariton chemistry community has harnessed ultrafast TA[70-75] and multidimensional[32, 76-79] techniques to probe polariton and dark state decay dynamics following direct optical pumping in systems under both electronic and vibrational strong coupling. Here, by contrast, we use ultrafast spectroscopy to track the photolysis-induced ground state chemical reactivity of molecules under VSC without any optical pumping of polariton states. The high IR reflectivity and visible transmission of our dichroic microfluidic cavities permits the interrogation of the dynamical signatures of CN radical reactions at wavelengths far from the strongly-coupled resonance, avoiding the cavity-induced TA artifacts that remain a significant topic of discussion in the literature.[79-81]

Our results demonstrate that the elementary CN + CHCl$_3$ reaction system shows negligible evidence of any reaction rate modification under VSC of the C−H vibrational mode of CHCl$_3$ when compared to experiments performed in off-resonance cavities and in standard IR-transparent flow cells. In as young and dynamic a field as polariton chemistry, examples of both significant *and* insignificant changes in reactivity under strong coupling are crucial to identify the conditions necessary to achieve cavity control of chemistry and validate or dismiss proposed mechanisms. We argue that the absence of cavity-altered reaction rates in this low-barrier, highly exothermic model system is consistent with recent literature findings[18, 38, 50, 82, 83] and supports the proposition that VSC may influence chemical reactivity by modulating vibrational energy transfer dynamics. In future follow-up experiments, we plan to systematically explore the impact of cavity-coupling strength and barrier height on elementary hydrogen abstraction reactions under VSC, providing further insight into the mechanisms of VSC-altered chemistry.

## II. EXPERIMENTAL METHODS

To assess the impact of VSC on elementary hydrogen abstraction reaction rates, we measure the dynamical signatures of CN radical reactions with CHCl$_3$ solvent molecules using ultrafast TA spectroscopy with a UV pump and visible white light (WL) probe. These experiments are performed within a microfluidic flow cell optical cavity tuned to strongly couple the C−H symmetric stretch of CHCl$_3$ located at 3020 cm$^{-1}$. We describe the TA apparatus utilized for these measurements in Section II.A and discuss the implementation of our dichroic cavity in Section II.B. We detail the preparation and handling of the reaction solutions in Section II.C.

### A. Ultrafast white light transient absorption spectroscopy

Experiments probing CN radical reaction dynamics under VSC require UV pump-WL probe measurements performed while monitoring the static IR spectrum of the cavity-coupled system. Our TA apparatus is shown in Fig. 3(a). A 7 mJ/pulse commercial Ti:sapphire femtosecond laser system (Astrella, Coherent) produces 60 fs pulses with a central wavelength of 800 nm at a 1 kHz repetition rate. The near-IR (NIR) output of the laser system is divided into three main arms by a series of beamsplitters to produce the UV pump, visible WL probe, and mid-IR monitor beams. A small fraction of the NIR output (~0.9 mJ) is additionally directed into a single-shot autocorrelator (Coherent) employed to monitor the pulse duration of the laser system (not shown in Fig. 3).

To generate the UV pump beam, we use an optical parametric amplifier (OPerA Solo OPA with FH/SHSF options, Light Conversion) to convert approximately 2.5 mJ of the NIR laser output into light with a central wavelength tunable between 240 – 2600 nm. All experiments described here employ OPA-generated pump pulses with a central wavelength of 250 nm (Fig. 3(b)). These 44 µJ/pulse pump pulses are attenuated before reaching the sample by a continuously variable neutral density filter (OD: 0.02 – 4.0, Thorlabs). Unless otherwise specified, the experiments were performed with the filter set to an optical density of 3.0, which was chosen to minimize cavity mirror artifacts and the accumulation of photoproducts in the relatively short pathlength sample cell, as described further in Section II.C.

A WL supercontinuum serves as the probe (Fig. 3(c)). For the probe beamline, we delay approximately 1.1 mJ of the NIR laser output relative to the UV pump using a motorized delay line (DL325, Newport) with a total of 325 mm (2.2 ns) of travel. We generate the WL supercontinuum by focusing the delayed NIR laser pulses with a 200 mm focal length lens into a 3 mm thick single crystal calcium fluoride flat (orientation [001], Eksma Optics).[84] The CaF$_2$ crystal is mounted on an actuated stage that continuously translates the crystal perpendicular to the pulse propagation direction to minimize laser-induced damage. To reduce periodicity due to the stage motion, we use an algorithm to programmatically select the next stage position according to a Gaussian distribution. The bandwidth and stability of the white light spectrum is optimized by modifying the input beam diameter and intensity with an iris and an attenuator composed of a $\lambda/2$ waveplate and thin film polarizer. Our resulting WL supercontinuum features spectral coverage spanning 320 – 500 nm and relative intensity fluctuations of less than 1.5%.

Both the UV pump and WL probe are focused into the sample cell by a 90° off-axis parabolic mirror (MPD249-F01, Thorlabs) with a reflected focal length of 101.6 mm. The pump beam is displaced relative to the probe beam on the parabolic mirror so that they intersect in the sample cell with a crossing angle of ~7°. The noncollinear beam geometry permits us to block the pump beam before it reaches the spectrometer, reducing residual pump light in the collected spectra. To eliminate the effect of anisotropic decay, a $\lambda/2$ waveplate located before the focusing optic sets the polarization of the pump beam to the magic angle (54.7°) relative to the probe beam.[85] At optimal spatial overlap, the pump beam features a beam diameter of ~1050 μm, roughly five times larger than the ~201 μm diameter of the probe beam.

Following recollimation, the WL probe is focused by a 100 mm focal length lens through a slit into a commercial Schmidt-Czerny-Turner spectrometer (Isoplane-320, Princeton Instruments). The WL is dispersed in frequency by a 300 groove/mm diffraction grating with a blaze wavelength of 300 nm (i3-030-300-P, Princeton Instruments) and recorded on a 1340 × 400 pixel CCD camera (Blaze-400B, Princeton Instruments). To limit dark current noise, the CCD camera is cooled to −80°C. All TA data is collected in a custom sensor mode that defines an active area of 1340 × 50 pixels to increase image acquisition speed and permit shot-to-shot measurements at 1 kHz. Data acquisition software programmed in LabVIEW (National Instruments) saves the spectrum of each shot impinging on the camera as a function of pump-probe delay. All subsequent data processing is performed in MATLAB and R Studio. Transient measurements are presented in absorbance ($A$):

$$A = -\log\left(\frac{I}{I_0}\right) \quad (1)$$

where $I$ and $I_0$ are the probe spectra transmitted through the sample in the presence and absence of the pump pulse, respectively. A mechanical chopper (MC2000, Thorlabs) modulates the pump at half the laser output repetition rate (500 Hz) such that an absorption spectrum can be constructed from each pair of laser pulses. All pulse-pair absorption spectra associated with a given pump-probe delay are averaged to generate a delay-dependent spectrum. To minimize the effect of long-term drift and improve the signal-to-noise ratio, we alternate the order in which pump-probe delays are collected between sequential datasets and average them. We then apply a three-component linear regression in R Studio and extract time constants following literature methods.[86] We find that chirp correction does not impact the measured time constants appreciably, but does introduce oscillatory artifacts in the spectra (see supplementary material, Section S1); we therefore do not apply chirp correction to the data presented here. We establish an instrument response function of ~120 fs using two-photon absorption signal in neat chloroform.[87]

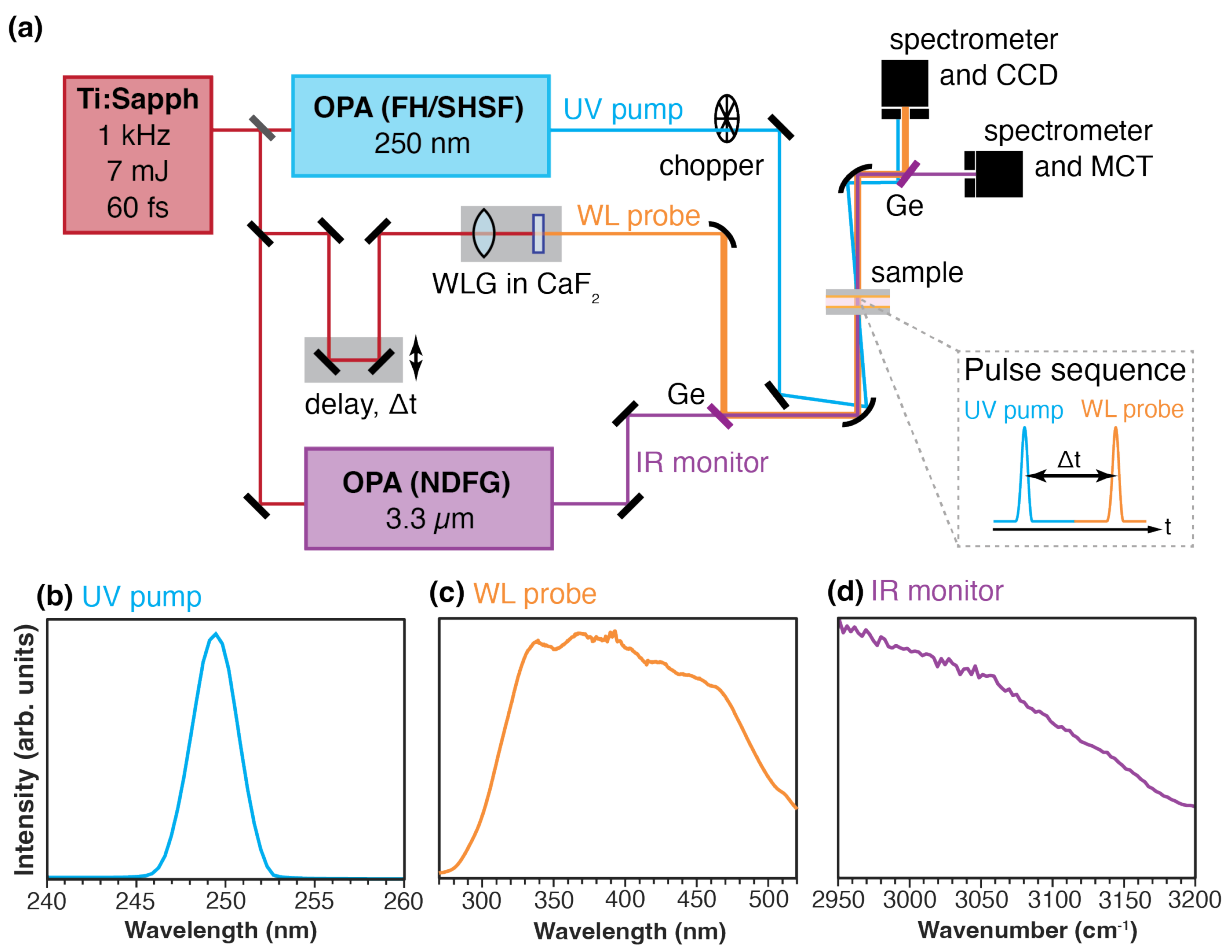

**FIG. 3.** Electronic transient absorption apparatus for measuring chemical reaction rates under vibrational strong coupling. **(a)** Our experimental setup permits both ultrafast ultraviolet (UV) pump − white light (WL) probe measurements and simultaneous monitoring of the static sample transmission at infrared wavelengths. **(b)** The UV spectrum of the 250 nm pump pulse. **(c)** The spectrum of the WL supercontinuum utilized as a probe in transient absorption experiments. **(d)** The infrared OPA spectrum covering the C−H stretching region.

We generate the mid-IR beamline for monitoring cavity transmission by directing the remaining ~2.5 mJ of the NIR laser output into a second optical parametric amplifier (OPerA Solo OPA with NDFG options, Light Conversion), producing light centered at 3400 nm (2940 cm$^{-1}$) with a full width at half maximum (FWHM) bandwidth of ~250 cm$^{-1}$. The 32 μJ/pulse mid-IR output passes through a germanium (Ge) window installed the probe beam path. The Ge window operates as a dichroic mirror reflective at visible wavelengths, permitting the transmitted IR pulses to propagate collinearly with the WL supercontinuum reflected off the surface. After focusing with a 90° off-axis parabolic mirror, the IR beam waist measures 300 μm in diameter in the sample cell. Spatial overlap between the IR, pump, and probe beams is verified by transmission through a 200 μm diameter pinhole. Following recollimation, the IR beam is transmitted through a second Ge window and focused by a 75 mm focal length lens into a commercial spectrometer (Acton SpectraPro, Princeton Instruments). The IR light is dispersed in frequency by a 75 groove/mm diffraction grating with a blaze wavelength of 4.6 μm and recorded by a 128 × 128 pixel Mercury Cadmium Telluride (MCT) focal plane array detector (2DMCT, PhaseTech). Array spectra with a resolution of ~2 cm$^{-1}$ are collected with PhaseTech QuickControl software, and all additional processing is performed in MATLAB. We calibrate the energy axis of the MCT detector by comparing measured spectra of neat chloroform and dichloromethane to reference spectra obtained with a commercial Fourier transform infrared (FTIR) spectrometer (Nicolet iS50 FT-IR, Thermo Scientific). A representative calibrated spectrum of our mid-IR light source is shown in Fig. 3(d). We collect IR cavity transmission spectra before and after TA measurements. During TA data collection, we block the IR beam immediately after the optical parametric amplifier to prevent optical pumping of the vibrational polaritons and undesired heating effects.

**B. Microfluidic dichroic Fabry-Pérot cavities for chemical reactions under VSC**

We install our microfluidic Fabry-Pérot cavities in a demountable flow cell with a 13 mm diameter aperture (TFC-M13-3, Harrick Scientific) designed for FTIR transmission measurements, following others in the VSC community.[73, 76] As shown in Fig. 4(a), the cavity consists of two mirrors separated by a polytetrafluoroethylene (PTFE) spacer (Harrick Scientific) that coarsely determines the cavity length ($L$). Spacers with a thickness of 56 µm provide a pathlength sufficient for absorption measurements while maintaining a cavity free spectral range (FSR) large enough for clear identification of polaritonic features. This ~100 cm$^{-1}$ FSR, in conjunction with the relatively simple IR spectrum of $CHCl_3$ in the C−H stretching region, helps ensure that only the target vibrational resonance couples to a single longitudinal cavity mode. We also performed experiments with thinner $L$~25 µm cavities, which exhibit a sparser cavity spectrum at the expense of TA signal (see supplementary material, Section S2). The cavity is sandwiched between two Viton O-rings that seal the liquid sample within the system when clamped in the demountable cell. We adjust the tightness of the cell to fine tune the cavity length and parallelism before performing spectroscopic measurements.

Key to the success of our experiments are custom-coated distributed Bragg reflector (DBR) mirrors (UltraFast Innovations GmbH) composed of alternating layers of high and low refractive index dielectric materials ($HfO_2$, $SiO_2$). These mirrors are designed for high reflectivity in the C−H stretching region of the IR spectrum to facilitate VSC while simultaneously maintaining transparency at UV and visible wavelengths to transmit the pulses we employ in TA measurements. Using a commercial IR microscope, we measure a mirror reflectivity of $R \geq 90\%$ from 2950 – 3365 cm$^{-1}$ and $R \sim 93\%$ at 3020 cm$^{-1}$ near the $CHCl_3$ C−H stretch targeted for VSC (see supplementary material, Section S3). The narrowband high-reflectivity window of our mirrors not only enables UV-Vis spectroscopy, but also prevents spurious cavity coupling of vibrational molecular resonances outside of the C−H stretching region.

When assembled into an air-filled 56 µm Fabry-Pérot cavity, the DBR mirrors transmit UV and visible wavelengths between 250 – 500 nm without any evidence of cavity mode structure or etaloning (Fig. 4(b)). In the C−H stretching region, the mirrors produce a cavity transmission spectrum characterized by ~6 cm$^{-1}$ FWHM cavity fringes spaced by a ~100 cm$^{-1}$ FSR, yielding an effective cavity finesse of $F \sim 17$ (Fig. 4(c)). We have designed the system to exhibit cavity fringe linewidths that closely match the 13 cm$^{-1}$ FWHM lineshape of the $CHCl_3$ C−H stretch, resulting in near-optimal conditions for polariton formation (see Section IIIA). The cavity fringes are slightly broader than expected given the measured mirror reflectivity, which we attribute to imperfect parallelism of the cavity mirrors. We exploit the imperfect parallelism of our cavity assembly to probe distinct cavity resonance conditions. Using actuated vertical and horizontal translation stages (Zaber), we tune the probed cavity length, and thus the cavity mode frequency, simply by translating the cell assembly in the plane of the cavity perpendicular to the propagation axis of the visible and IR beams.

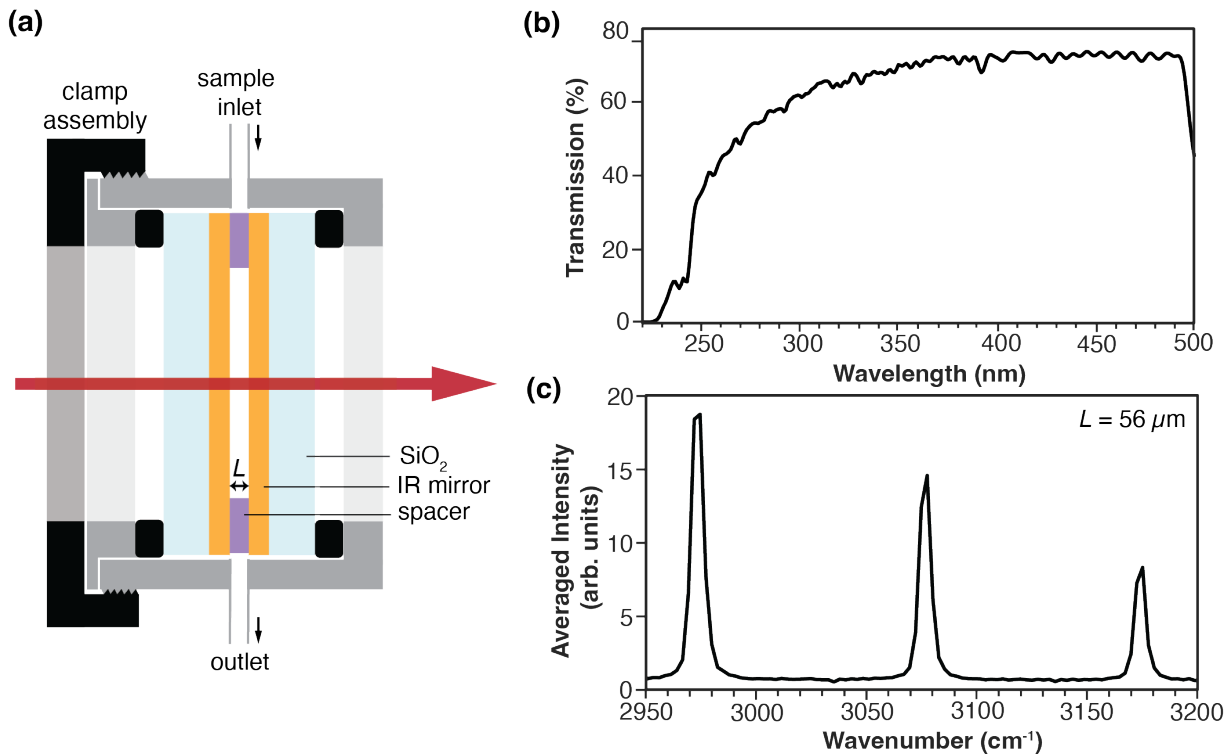

**FIG. 4.** Dichroic microfluidic cavity used for ultrafast transient absorption measurements performed under VSC. **(a)** A schematic of a Fabry-Pérot cavity composed of IR-reflective DBR mirrors integrated within a microfluidic flow cell. Experimental **(b)** UV-Vis transmission spectrum for pump and probe wavelengths and **(c)** IR cavity transmission spectrum of the C−H stretching region obtained from an empty cavity constructed with a 56 μm spacer.

For TA measurements, we monitor the cavity transmission spectrum while translating the cell until we achieve the desired coupling condition. We record the IR transmission spectrum before blocking the IR beam to prevent photonic excitation of the vibrational polariton and undesired heating of the sample. After the collection of each TA dataset, we record a second IR spectrum to verify that the cavity coupling condition did not change during the experiment. We then translate the cell to a fresh, unused position to avoid the accumulation of opaque photoproducts on the surface of the mirrors.

Throughout this work, we compare intracavity results for various detuning conditions to *extracavity* control measurements performed in an identical microfluidic cell fitted with IR-transparent $CaF_2$ windows and a 56 μm spacer. We use the extracavity controls to validate that our data collection and processing protocols yield measurements consistent with prior literature.

## C. ICN sample preparation and handling

We prepare 0.3, 0.43, and 0.5 M ICN solutions by dissolving ICN (98% purity, Thermo Scientific Chemicals) into neat liquid chloroform (99.8% purity, Sigma-Aldrich) without any additional purification.[62] These solutions are delivered to the microfluidic sample cell at a rate of 12 – 17 mL/min by a peristaltic pump (Masterflex) through Viton tubing with an inner diameter of 0.80 mm (Avantor). Notches cut in the cavity spacer adjacent to the inlet and outlet of the microfluidic cell permit the unimpeded flow of liquid sample through the cell. The continuous flow of solution through the cell during transient absorption measurements reduces the accumulation of unwanted photoproducts and dissipates heat generated by the UV pump pulse and the exothermic reaction. To ensure that the cell assembly is at thermal equilibrium during experiments, we measure the temperature of the demountable flow cell assembly after being placed on the warm actuated stages using a thermocouple fastened within a port intruding into the body of the cell. For both empty and $CHCl_3$-containing cells, the temperature rises by > 2°C before stabilizing over a period of 90 minutes. We therefore allow the cell to thermally equilibrate for at least 90 minutes before performing measurements.

## III. RESULTS
### A. Vibrational strong coupling of the C−H stretching mode in CHCl₃ solvent

We first demonstrate that we can reach the strong coupling regime for neat chloroform in our DBR optical microcavities. The C−H stretching vibrational mode of chloroform appears as an isolated, asymmetric resonance at 3020 cm$^{-1}$, as shown by the experimental absorption spectrum in Fig. 5(a)). The linewidth of this mode (~14 cm$^{-1}$ FWHM) is slightly broadened relative to literature values[88] due to the ~2 cm$^{-1}$ resolution of our spectrometer. To engineer VSC, we fit the microfluidic cell with DBR mirrors and inject neat $CHCl_3$. We construct a dispersion curve by recording the cavity transmission spectrum as a function of cavity length. A representative dispersion curve is shown in Fig. 5(b). We convert the translation axis to cavity length by fitting the spectra to the classical expression for light transmitted through a Fabry-Perot cavity (*vide infra*).

As expected, our experimental cavity dispersion spectra feature an avoided crossing as the cavity fringe passes through resonance with the chloroform C−H stretch, yielding a measured Rabi splitting of 25 cm$^{-1}$. This Rabi splitting exceeds both the molecular (~14 cm$^{-1}$ FWHM) and cavity (12 cm$^{-1}$ FWHM) linewidths, indicating that the system has reached the VSC regime. In Fig. 5(c), we present cavity transmission spectra depicting representative on- and off-resonance cavity tuning conditions that we employ in TA experiments (see Section III.B).

Here, we designate the on-resonance cavity tuning condition as the case where the upper and lower polariton peaks exhibit roughly equal transmission amplitudes. We additionally identify an alternative near-resonance cavity tuning condition when the splitting between upper and lower polariton features is minimized; this "minimum Rabi splitting" case features an upper polariton which is less intense than the lower polariton due to the asymmetry of the $CHCl_3$ absorption feature. Cavity transmission spectra obtained at the minimum Rabi splitting coupling condition

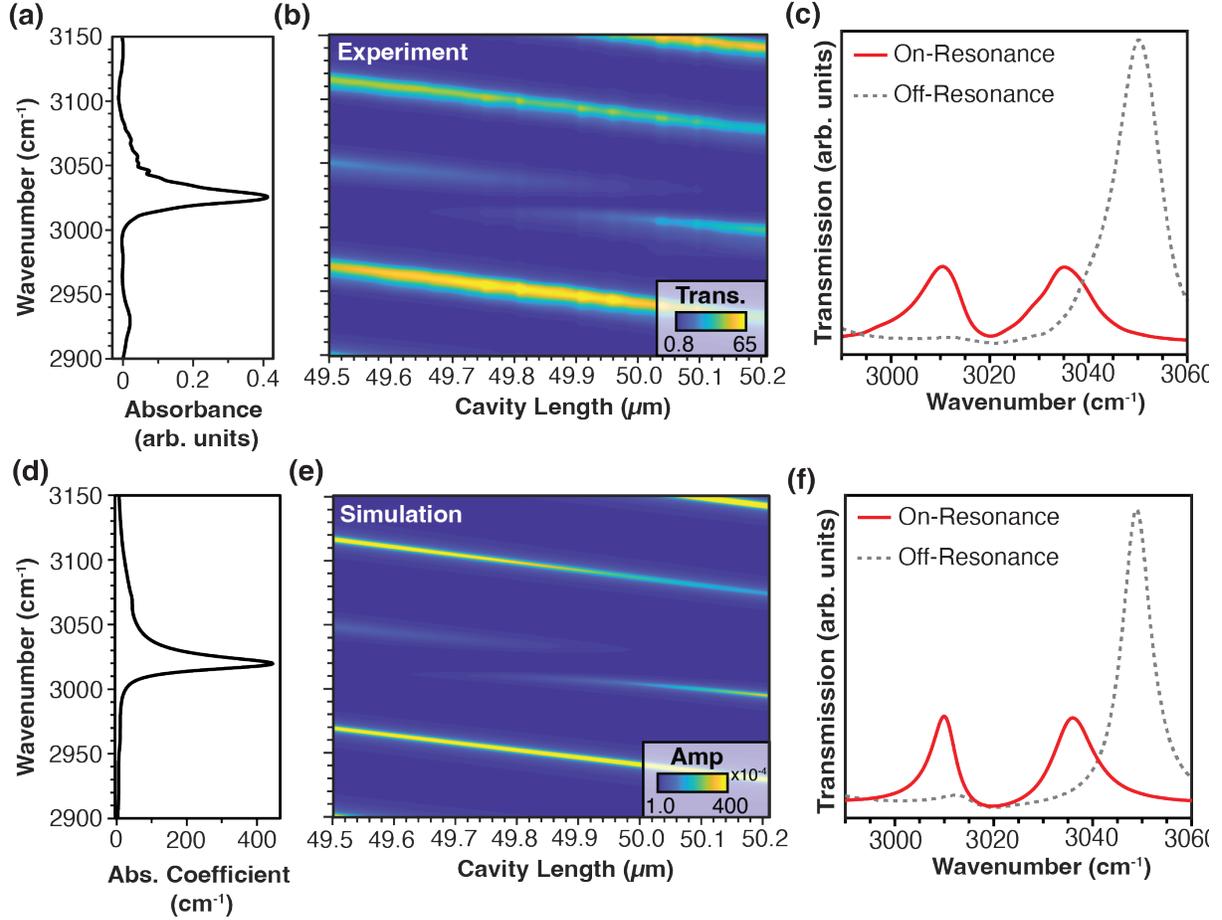

**FIG. 5.** Experimental and simulated absorption spectra and cavity transmission spectra for neat $CHCl_3$ in a DBR Fabry-Pérot cavity with a spacer of nominal length $L=56$ μm. **(a)** Experimental absorption spectrum of the $CHCl_3$ C−H stretching mode measured in a cell fitted with IR-transparent windows. **(b)** Experimental cavity transmission spectra of neat $CHCl_3$ measured as a function of cavity length. **(c)** Experimental transmission spectra representative of cavities tuned on- and off-resonance with the $CHCl_3$ C−H stretch. **(d)** Absorption coefficient of the $CHCl_3$ C−H stretching mode calculated from refractive index data provided by PNNL. **(e)** Simulated transmission spectra of intracavity $CHCl_3$ plotted as function of cavity length using Eqn. 2. **(f)** Simulated cavity transmission spectra for cavity lengths tuned on- and off-resonance with the $CHCl_3$ C−H stretch.

are provided in supplementary material, Section S4.

We simulate our experimental cavity transmission spectra using the classical expression for the relative intensity of light transmitted through a Fabry-Pérot cavity composed of two identical mirrors:[89-91]

$$\frac{I_T(\nu)}{I_0} = \frac{T(\nu)^2 e^{-\alpha(\nu)L}}{1+R(\nu)^2 e^{-2\alpha(\nu)L} - 2R(\nu)e^{-\alpha(\nu)L}\cos\left(\frac{4\pi L n(\nu)\nu}{c}\right)} \quad (2)$$

Here, $T(\nu)$ and $R(\nu)$ are the frequency-dependent transmission and reflectivity coefficients for a single DBR mirror, $\alpha(\nu)$ and $n(\nu)$ are the absorption coefficient and real component of the refractive index of the intracavity material, and $L$ is the cavity length. Reference data for the real ($n(\nu)$) and imaginary ($k(\nu)$) components of the complex refractive index were provided by Pacific Northwest National Laboratory (PNNL).[88] We convert $k(\nu)$ to absorption coefficient $\alpha(\nu)$ using the following expression:[92]

$$\alpha(\nu) = 4\pi k(\nu)\nu \quad (3)$$

We plot the frequency-dependent absorption coefficient of chloroform in the C−H stretching region in Fig. 5(d), which features a single resonance with a 13 cm$^{-1}$ FWHM linewidth.

In Fig. 5e, we simulate the transmission spectrum of a CHCl$_3$-containing cavity as a function of cavity length using Eqn. 2. The simulated dispersion spectra exhibit an avoided crossing indicative of strong light-matter coupling of the C−H stretch of CHCl$_3$. The minimum difference in frequency between the upper and lower polariton features, or the Rabi splitting, is calculated to be 26 cm$^{-1}$, exceeding the linewidths of both the molecular (13 cm$^{-1}$ FWHM) and simulated cavity (3 cm$^{-1}$ FWHM) resonances. This simulated Rabi splitting is in nearly perfect agreement with our experimentally measured Rabi splitting of 25 cm$^{-1}$. This agreement is further emphasized by comparing simulated cavity transmission spectra at specific on- and off-resonance tuning conditions (Fig. 5(f)) with their experimental counterparts (Fig. 5(c)). Any discrepancies between the experimental and simulated spectra arise primarily from differences in cavity mode linewidths, which are broader experimentally due to finite spectrometer resolution and the nonuniformity of cavity lengths in the experimental IR focal volume.

**B. Transient absorption measurements of intracavity CN radical reaction dynamics**

Our dichroic cavity architecture enables direct spectroscopic detection of the electronic signatures of CN radicals as they react with strongly-coupled CHCl$_3$ solvent. Following UV photolysis of ICN, we monitor the ultrafast evolution of the CN radical population via WL TA spectroscopy. To assess the impact of VSC on CN radical reaction dynamics, we measure broadband transient spectra under extracavity, on-resonance, and off-resonance coupling conditions. The resonance conditions corresponding with DBR cavities tuned either on- or off-resonance with the C-H stretching mode of CHCl$_3$ are illustrated in Fig. 5(c,f). An alternative near-resonance condition is presented in the supplementary material, Section S4. We verify the cavity coupling conditions before and after all TA measurements. Because the IR pulse is blocked during data collection, the CN radical rate constants reported here are not impacted by photonic excitation of the vibrational mode or IR-induced heating effects.

We analyze the TA spectra obtained for the CN + CHCl$_3$ system using well-established procedures deployed in previously published ICN photodissociation experiments.[63] Fig. 6(a) shows a representative false color plot of the absorption signal obtained from a 0.5 M ICN solution in CHCl$_3$ as a function of wavelength and pump-probe delay in an on-resonance cavity of nominal length $L$ = 56 μm. The two broad, positive transient absorption features centered at 390 nm and 340 nm overlap spectrally, but exhibit distinct dynamics in the time domain. In order to extract meaningful time constants from these overlapping features, we implement a multiple linear regression to decompose the absorption spectrum acquired at each pump-probe delay into three spectral components. A representative decomposition of a TA spectrum measured at a pump-probe delay of 1.4 ps is shown in Fig. 6(b). The regression captures the free CN radical feature at 390 nm with a Lorentzian function fit to a spectrum taken at an early time delay (1.2 ps) before the emergence of other complicating features. We model the solvated CN radical feature at 340 nm using a late time delay spectrum (1000 ps) obtained after the 390 nm feature has decayed. The final component of the multiple linear regression captures a broad I*($^2P_{1/2}$)-solvent charge transfer band centered at 540 nm,[63] and accounts for UV-induced mirror and pure solvent TA features. The dielectric coatings of the DBR mirrors absorb a fraction of the pump light, generating a broad TA signal that persists beyond the 1.9 ns time delay window interrogated in these experiment (see supplementary materials, Section S5). Moreover, the absorption of two pump photons by neat

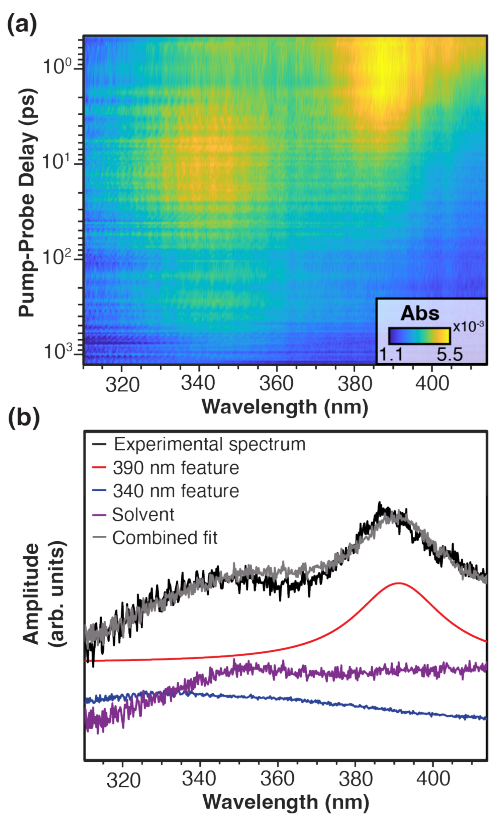

**FIG. 6.** Transient absorption of UV photolysis-initiated dynamics in an 0.5 M ICN solution in CHCl$_3$ as measured through an on-resonance infrared DBR microcavity of nominal cavity length $L$=56 μm. **(a)** Representative electronic transient absorption spectrum plotted as a function of wavelength and pump-probe delay. Note that the vertical pump-probe delay axis is plotted on a logarithmic scale to better depict dynamics occurring on disparate timescales. Warmer colors correspond to the appearance of transient features associated with the formation of new species following photolysis. **(b)** A representative spectral decomposition for an experimental spectrum (black) obtained at a pump-probe delay of 1.4 ps. The 390 nm (red), 340 nm (blue), and solvent (purple) features are modelled respectively by a Lorentzian function, the transient probe spectrum at a 1000 ps pump-probe delay, and a neat intracavity CHCl$_3$ absorption spectrum obtained at 100 ps. The components sum to the combined fit (grey).

CHCl$_3$ solvent results in a photoproduct that continues to absorb long after the dissipation of the signal used to determine pump-probe time overlap.[87] Collectively, these processes result in a broad Gaussian-like feature with a bandwidth broader than our probed spectral window. For simplicity, we use the broad absorption measured in a DBR cavity containing neat CHCl$_3$ at 100 ps as the input for the third component in the regression and refer to it as the "solvent" feature. We plot the amplitude of each of these regression components as a function of pump-probe time delay to recover time traces for the 390 and 340 nm features.

To establish mechanistic assignments for the observed spectral signatures, we compare our measured extracavity results to the existing literature in Table I. Our extracavity measurements are detailed further in supplementary material, Section S1. UV photolysis of molecular ICN is known to generate free CN radicals identifiable by their absorption signature at 390 nm due to the CN $\tilde{B} \leftarrow \tilde{X}$ electronic transition.[62, 68] Following its rapid appearance at early time delays, the free CN radical population decays via multiple pathways (Fig. 2(b)). The contribution of each of these pathways to the evolution of the 390 nm feature remains an active point of discussion, but most prior work agrees that its temporal profile is best fit to a biexponential decay.[61, 62, 69] In the literature, the initial $\tau_1 \sim 4$ ps decay of the 390 nm feature is ascribed predominantly to primary geminate recombination of the I and CN fragments and isomerization to INC within the solvent cage,[61, 62, 67, 69] as well as to the formation of CN-solvent complexes.[63, 64] The value and origin of the longer 390 nm decay constant is less clear. While Crowther et al.[69] reports a long time constant of $\tau_2 \sim 1500$ ps, Wan et al.[61] and Moskun and Bradforth[62] agree that $\tau_2 \sim 70$ ps. This slow decay is typically attributed to CN reaction with solvent, including hydrogen abstraction,[61, 62, 69] and secondary diffusive recombination after cage escape.[62] Here, we measure extracavity decay rates of $\tau_1 = 4.1 \pm 0.3$ ps and $\tau_2 = 70 \pm 6$ ps for the 390 nm feature, in excellent agreement with prior work (Table I). As the 390 nm feature decays, a broader, blue-shifted feature centered near 340 nm emerges. Orr-Ewing and coworkers report that this 340 nm feature grows in with a time constant of $\tau_1 \sim 2$ ps before evidencing a much slower $\tau_2 \sim 1500$ ps decay.[63] They attribute the growth of the 340 nm feature to the formation of CN-solvent complexes, with additional contributions from an underlying I($^2P_{3/2}$)-solvent charge transfer band. The subsequent decay of the 340 nm feature can be explained both by hydrogen abstraction and by the diffusive recombination of CN radicals with I*($^2P_{1/2}$) atoms quenched to the I($^2P_{3/2}$) ground state.[63] In this work, we measure an extracavity growth rate of $\tau_1 = 1.7 \pm 0.7$ ps and a decay rate of $\tau_2 = 1500 \pm 150$ ps for the 340 nm feature (Table I). We therefore establish an extracavity baseline for this system consistent with prior work, and can proceed to screen for cavity modification of these measured time constants.

In Fig. 7, we show representative time traces obtained from TA datasets taken in DBR cavities of nominal length $L$=56 μm tuned on- and off-resonance with the CHCl$_3$ stretch. As detailed in Section II.B, we tune the cavity length, and thus the mode frequency, by translating the cell in the plane of the cavity. In cavities tuned off-resonance with the CHCl$_3$ C−H stretch, the evolution of the 390 nm feature remains well-fit to a biexponential decay, again due to the loss of the free CN population to recombination, isomerization, and solvent-complexation processes as in the extracavity system (Fig. 7(a)). The 340 nm feature also remains well-described by fast exponential growth followed by a slow decay in off-resonance cavities, as solvent-complexed CN species are rapidly formed and then slowly lost to hydrogen abstraction reactions with solvent (Fig. 7(b)). The time traces of the 390 nm and 340 nm features measured in cavities tuned *on*-resonance with the CHCl$_3$ C−H stretch (Fig. 7(c) and 7(d)) do not differ appreciably from their off-resonance counterparts within our temporal resolution. Table I compares the average 340 nm and 390 nm time constants calculated from 16 pairs of on- and off-resonance datasets taken over a period of four months. More details on this data set are provided in supplementary material, Section S6. Time constants extracted from TA measurements performed at the alternative near-resonance cavity coupling condition are provided in supplementary material, Section S7. All time constants obtained under on-resonance strong coupling conditions differ by less than one standard deviation from those obtained in off-resonance cavities, and are similar to the rate constants procured from our extracavity measurements (Table I). Moreover, these intracavity time constants are consistent with extracavity measurements reported in the literature for both the free CN radical feature at 390 nm[61, 62] and the solvated CN feature at 340 nm.[63] These results indicate that VSC of the C−H stretching mode of $CHCl_3$ has a negligible impact on the recombination, isomerization, solvent-complexation, or hydrogen abstraction dynamics of CN radicals.

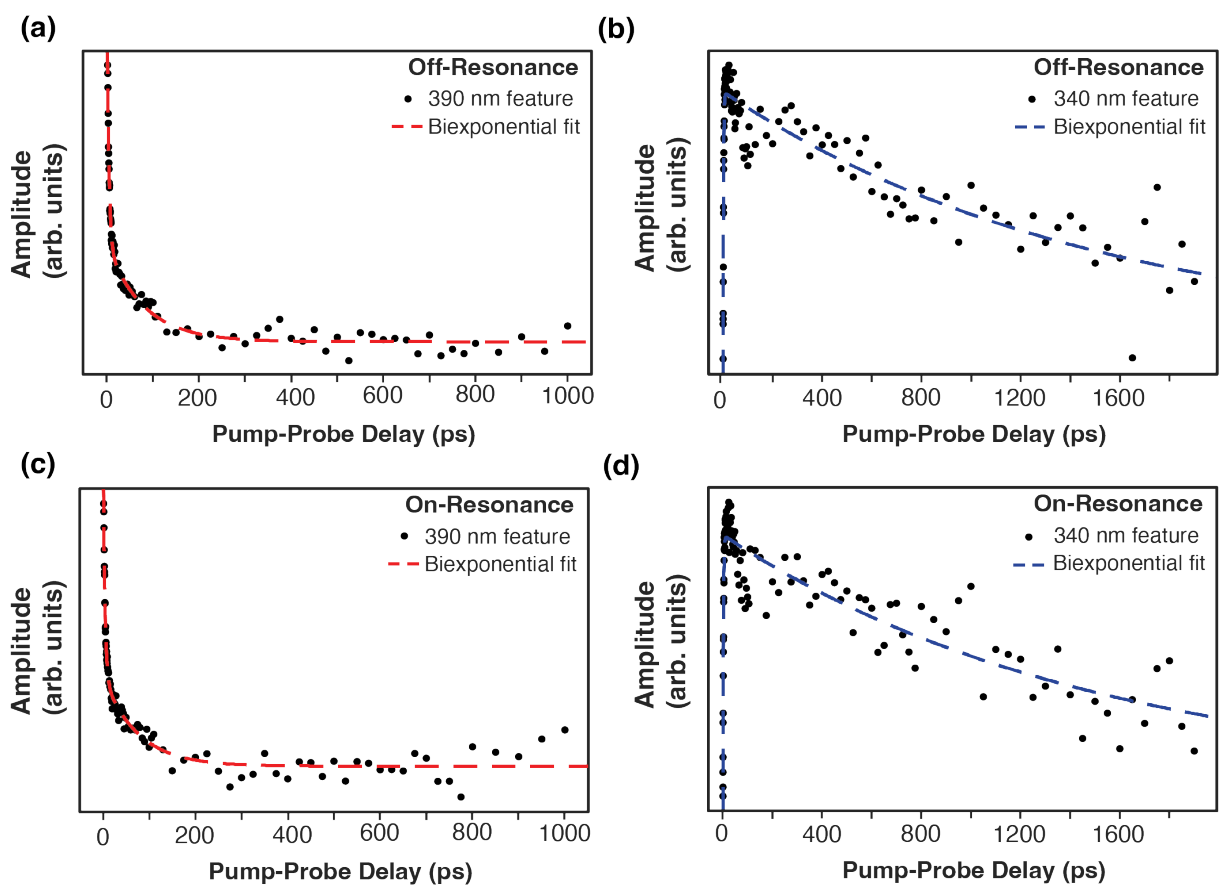

**FIG. 7.** Representative intracavity CN radical reaction dynamics measured under different coupling conditions in DBR cavities. Time traces and biexponential fits of the **(a)** 390 nm and **(b)** 340 nm spectral components extracted from TA measurements when the cavity is tuned off-resonance with the C−H stretch of CHCl$_3$. Similar traces and fits are obtained for the **(c)** 390 nm and **(d)** 340 nm features under on-resonance conditions. Representative cavity transmission spectra for on- and off-resonance cavity coupling conditions are shown in Fig. 5(c,f).

TABLE I. Time constants for the CN + $CHCl_3$ reaction system measured in cavities tuned on- and off-resonance with the C−H symmetric stretch of $CHCl_3$, as compared to extracavity results performed in this work and in prior literature.

| **390 nm** | On-resonance (this work) | Off-resonance (this work) | Extracavity (this work) | Orr-Ewing[63] | Bradforth[62] | Crim[69] | Zewail[61] |
|---|---|---|---|---|---|---|---|
| $\tau_1$ (ps) | 3.4 ± 0.5 | 3.2 ± 0.6 | 4.1 ± 0.3 | 3.3 ± 1.0 | 4 | 4 | 2.5 ± 0.4 |
| $\tau_2$ (ps) | 70 ± 15 | 71 ± 16 | 70 ± 6 | - | 72 | 1500 | 70 ± 10 |
| **340 nm** | On-resonance (this work) | Off-resonance (this work) | Extracavity (this work) | Orr-Ewing[63] | Bradforth[62] | Crim[69] | Zewail[61] |
| $\tau_1$ (ps) | 2.3 ± 0.6 | 2.4 ± 0.6 | 2.5 ± 0.2 | 1.7 ± 0.7 | – | – | – |
| $\tau_2$ (ps) | 1500 ± 400 | 1600 ± 400 | 1420 ± 40 | 1500 ± 150 | – | – | – |

## IV. DISCUSSION

Here, we have developed the infrastructure to explore polariton chemistry in a new experimental regime that may better elucidate the mechanisms responsible for VSC-altered reactivity. As well-characterized elementary processes with clear electronic transient signatures, photolysis-initiated CN radical reactions represent a compelling model system by which to directly track ultrafast intracavity molecular dynamics and bimolecular reaction kinetics. We do not observe significant modification to any reaction rate constants of the CN + $CHCl_3$ reaction system under VSC of the C−H stretching mode of $CHCl_3$. In Section IV.A, we detail the procedures we have developed to ensure the robustness of our intracavity results. In Section IV.B, we propose potential explanations for our observation of negligible cavity modification of ground state chemistry in the CN + $CHCl_3$ reaction system.

### A. Practical experimental considerations for probing elementary reactions under VSC

Here, we address the challenges encountered while examining this ultrafast elementary reaction system into a cavity architecture designed to facilitate VSC, including the necessity for an IR beamline to monitor cavity transmission, secondary photoproduct accumulation, and spurious TA signals from our DBR mirrors.

#### *1. Monitoring the cavity coupling conditions in situ*

Due to the imperfect parallelism of our microcavities, measurements performed on a separate instrument, such as a commercial FTIR spectrometer, cannot provide an accurate measurement of the cavity coupling conditions in the exact spot probed by a given TA measurement. We therefore found it essential to monitor the cavity transmission spectrum *in situ* using IR light spatially overlapped with the white light probe. We record the cavity transmission spectrum before and after the collection of each TA dataset to confirm that the desired cavity-coupling conditions are maintained throughout the course of the experiment. The IR pulse is blocked during TA experiments to prevent any undesired heating or optical pumping effects. Instability in the cavity length, and consequentially the cavity transmission spectrum, arises primarily from the initial introduction of solution to the cell and warming from the raster stages. In future work, active cavity stabilization using a piezoelectric element may prove beneficial, particularly if rastering the cavity during experiments is deemed necessary (*vide infra*).

*2. Accumulation of secondary reaction products on cavity mirrors*

An opaque photoproduct from secondary radical reaction products accumulates on the windows during our TA measurements. These deposits gradually reduce transmission through the microfluidic system, resulting in lower quality TA signals and deterioration of the cavity finesse over time. Photoproduct buildup on window surfaces has been reported in other spectroscopic measurements, motivating the development of rastering schemes that continuously translate the cell[93, 94] and liquid sample jets that eliminate the need for windows entirely.[95, 96] To minimize photoproduct accumulation, we successfully implemented a rastering scheme for our extracavity measurements. However, when making measurements in DBR cavities with imperfect mirror parallelism, translation of the cell modifies the cavity-coupling conditions, making it difficult to correlate TA signals with cavity resonance conditions while rastering. Instead, we optimized our data acquisition parameters, including the velocity of the delay line and the readout time of the CCD camera, to reduce the collection time for each dataset while maintaining a workable signal-to-noise ratio upon averaging. In addition, we instituted measures to counter the impact of the long-term attenuation of transmitted signal by the opaque photoproduct. Critically, the absorbance of the CN radical features remains roughly constant over time even as the probe transmission drops because we collect pump-off reference spectra every other laser pulse. We also vary the order in which we collect pump-probe time delays, preventing deterioration of the TA spectra from appearing as an additional delay-dependent decay in our time-domain data.

*3. Transient absorption artifacts in the DBR mirrors*

Another important consideration in our experiments are the spurious transient absorption signals arising from the DBR cavity mirrors. Significant UV absorption (Figure 4(b)) by the high-*n* $HfO_2$ layers of the dielectric mirrors results in broadband TA signals following pump excitation (see supplementary material, Section S5). Due to the small optical pathlength through the cavity, TA signals from both the mirrors and the target ICN solution contribute to our data. We cannot readily replace the DBR mirrors with metallic mirrors like those commonly employed in the VSC literature[37] because we have found that the ultrafast UV and WL laser pulses used in these experiments ablate thin gold and silver coatings. As described in Section III.B, we include the TA signal from the DBR mirrors in the spectral decomposition of our data to minimize its impact on the extracted time constants.

**B. Explanation of negligible cavity-altered chemistry in CN + CHCl₃ reactions**

A substantial body of literature has reported VSC-altered chemical kinetics in relatively slow, complex chemical reactions with high activation barriers.[18, 39, 40, 43, 44] We now discuss how the negligible evidence of rate modification in our ultrafast, elementary reaction system may be attributed to a number of potential causes.

*1. Limitations in collective coupling strength*

The collective cavity coupling strength achieved in our experiments may be too small to measurably change reaction rates. While chloroform is an attractive solvent for these measurements because of its simple IR spectrum and well-characterized TA signals, the 25 cm$^{-1}$ Rabi splitting we obtain by strong coupling the C−H symmetric stretch in neat CHCl₃ is conservative compared to the >70 cm$^{-1}$ splittings reported for reaction systems in the literature.[18] A direct comparison between our experimental parameters and those reported in the polariton chemistry literature is provided in supplementary material, Section S8. We can increase the collective coupling strength by choosing a solvent system with a larger IR transition dipole than can be found in CHCl₃. CN radical reaction dynamics have been previously studied in a wide array of solvent systems,[63, 65, 66, 97] many of which possess far brighter C−H stretching transitions than CHCl₃. For example, targeting a cyclohexane-chloroform solvent system would allow us to achieve Rabi splittings not only larger than that observed in CHCl₃, but also tunable via the relative fraction of each co-solvent. We will explore this broader class of CN-solvent hydrogen abstraction reactions in future work.

In addition, we may be able to enhance polaritonic effects in our system by decreasing the cavity mode volume in order to increase the effective collective coupling strength per molecule. Our measurements in nominally $L$ = 56 μm cavities make use of longitudinal cavity modes of higher mode order ($m \sim 40-50$) than are typically used in literature VSC chemistry experiments.[39, 41, 43, 44] Previous literature results have demonstrated that the collective Rabi splitting scales only with intracavity absorber concentration according to $(N/V)^{1/2}$ and is independent of cavity length.[98, 99] However, the impact of cavity mode volume on VSC-modified chemistry has not been fully characterized.[91] Reducing our cavity lengths from 56 μm to 25 μm did not appreciably change the measured Rabi splitting in our system (see supplementary material, Section S2), but did reduce the quality of our TA signal. In future work, we may improve our spectroscopic infrastructure in order to perform TA measurements in thinner cavities with higher effective cavity coupling strengths per molecule. Improvements to the design of the DBR cavity mirrors represent an alternative route toward enhanced coupling strength, as mode penetration into the dielectric layers can result in fewer coupled molecules within a given mode volume.[98]

*2. Obscured signatures of cavity-altered dynamics*

Evidence of any VSC-altered chemistry in our system could be overwhelmed by transient signatures of either CN radical populations reacting with uncoupled $CHCl_3$ molecules or dynamical processes unaffected by VSC, such as geminate recombination. As in many polariton chemistry experiments, we expect that our measurements inherently include background signals from reactions of uncoupled molecules. In our system, the uncoupled $CHCl_3$ reactant population includes molecules whose C−H stretching transition dipole is oriented perpendicular to the plane of the cavity mirrors, as well as molecules not located at the field maxima of the standing wave that defines the coupled longitudinal IR cavity mode.[12] IR cavity transmission spectroscopy is conveniently blind to these uncoupled molecules, as the molecules that contribute to optical absorption signals are the same population that experiences strong cavity coupling.[81] In our experiments, however, our visible WL probe is not confined by the infrared cavity and will encode absorption signals from both cavity-coupled and uncoupled molecules. Given these challenges, cavity-modified dynamics from minor reaction pathways, such as hydrogen abstraction, may be obfuscated in our TA data.

We should also note that the dynamics in our TA spectra are dominated by the ~85% of CN radicals that undergo recombination with atomic iodide, a process unlikely to be perturbed by VSC of the surrounding solvent.[58] We might obtain more information about the time constants associated with target pathways, particularly hydrogen abstraction, by monitoring the formation and vibrational distribution of the products explicitly, as has been done in the extracavity literature.[57, 63, 65] To avoid artifacts from cavity spectral filtering,[79-81] it is desirable to perform spectroscopic measurements of reaction dynamics at wavelengths far removed from the polariton states. Probing product formation either directly in a region of the IR spectrum outside the DBR high-reflectivity band or at optical wavelengths via time-resolved Raman spectroscopy may therefore prove beneficial in isolating pathways potentially impacted by VSC.

### 3. Dynamical mechanisms for vibrational polariton chemistry

Finally, we consider the possibility that the mechanisms by which vibrational polaritons impact chemistry simply do not act on ultrafast, low-barrier reactive processes. The polariton chemistry community is increasingly reaching the consensus that VSC influences chemical reactions by altering vibrational dynamics, rather than by reshaping reactive potential energy surfaces.[16-19] In particular, a growing body of theoretical and experimental work implicates cavity-mediated IVR channels as one key mechanism by which VSC may operate on reactive trajectories.[18, 50-54, 82] In one compelling experimental example, Chen *et al*.[82] observe that optically pumped vibrational polaritons in $Fe(CO)_5$ have a higher propensity to decay via IVR rather than undergo unimolecular pseudorotation when compared to the corresponding bare molecular vibrations.[19, 100]

The systems in which *thermal* VSC-altered kinetics have been reported feature slow processes between large molecules with many vibrational degrees of freedom. These reactions proceed over relatively large activation barriers, implicating the involvement of thermal trajectories that access highly excited vibrational states. The most compelling examples of VSC-altered chemistry report rate reductions,[39, 43, 101] potentially because cavity-mediated IVR processes drain energy away from the reaction coordinate. In contrast, the highly exothermic CN + CHCl$_3$ hydrogen abstraction system we examine here possesses a negligible activation barrier that is rapidly surmounted at room temperature without high levels of reactant thermal vibrational excitation. As this reaction can proceed in its ground vibrational state, we should perhaps not expect that the CHCl$_3$ vibrational polaritons would dominate the reactive dynamics, nor that cavity-mediated IVR pathways would significantly perturb the reaction rate. Thus, our report of the negligible impacts of VSC on CN + CHCl$_3$ reaction dynamics is consistent with the growing consensus that cavity-mediated redistribution of vibrational energy may be an important mechanism in polariton chemistry.

At present, it appears that VSC has the most pronounced impact on the chemistry of complex, high-dimensional solution-phase reactions, in stark contrast to the traditional domains of mode-selective chemistry and coherent control. While vibrational mode-selective chemistry has historically been restricted to simple reaction systems where IVR does not outcompete the target reaction,[2, 3, 11] polariton chemistry may in fact *rely* on the dominance of IVR or related vibrational energy transfer mechanisms to steer reactive trajectories. If so, polaritonic platforms may indeed permit the extension of photonic control schemes to solution-phase chemistry, realizing a long-held ambition of physical chemistry. One future avenue may be to move on from the low-barrier reaction regime studied here towards systems with higher barriers in which vibrational dynamics are more likely to dominate reactivity and thus may be more susceptible to perturbation by VSC. Drawing inspiration from the vibrationally-driven chemistry literature, one might consider strong coupling of *endothermic* bimolecular reactions that only proceed following the injection of energy by vibrational pumping.[102] In such a system, one could directly compare the relative reactive propensities of pumped bare molecular vibrational states versus pumped vibrational polaritons, without relying on thermal occupation of vibrational states. In addition, the recent demonstration of gas-phase molecular polaritons in our laboratory[103] will provide another means to examine how simple reaction systems from the mode-selective chemistry literature might behave under VSC.

## V. CONCLUSIONS

Vibrational polariton chemistry, while still in its infancy, has great potential as a new toolkit for rational control of chemical reactions. Here, we have set out to elucidate the mechanisms underlying cavity-altered chemical reactivity by directly tracking the dynamics of a benchmark bimolecular solution-phase reaction under VSC, specifically targeting the low-barrier elementary CN + CHCl$_3$ reaction. Using ultrafast electronic transient absorption spectroscopy, we observe no change in reaction rate under VSC of the C−H stretching mode of CHCl$_3$. Our work represents the first reported study of laser-induced ultrafast bimolecular reaction dynamics under vibrational strong coupling. The conspicuous absence of cavity-altered dynamics in this reaction system serves as a new data point regarding the effects of strong coupling on chemical reactivity in a regime completely distinct from prior work. We believe that null results are essential data points for validating proposed mechanisms of cavity chemistry, and we join others in the field in urging experimentalists to publish results in which VSC fails to alter reactivity.[17, 45-47] Through continued conversation with theory, our work will help leverage molecular polaritonics for control of chemical reactions and inform prospects for using the new degrees of freedom afforded by cavity coupling to direct increasingly complex chemistry.

## SUPPLEMENTARY MATERIAL

See the supplementary material for additional characterization of the DBR cavity mirrors, a detailed breakdown of the transient absorption measurements averaged to obtain the overall rate constants reported in this work, and transient absorption data obtained under an alternative cavity resonance condition.

## ACKNOWLEDGMENTS

This work was supported by the US Department of Energy, Office of Science, Basic Energy Sciences, CPIMS Program under Early Career Research Program award DE-SC0022948 and by Princeton University start-up funds. We also acknowledge funds from the Princeton Center for Complex Materials (PCCM) National Science Foundation (NSF) Materials Research Science and Engineering Center (MRSEC; DMR-2011750), which supported the construction of our infrared spectrometer. APF acknowledges support from the Princeton Presidential Postdoctoral Research Fellowship. We thank Tanya Myers for providing quantitative frequency-dependent complex refractive index data for chloroform measured by Pacific Northwest National Laboratory (PNNL) and Intelligence Advanced Research Projects Activity (IARPA).

## AUTHOR DECLARATIONS

**Conflict of Interest**

The authors have no conflicts to disclose.

**Author Contributions**

**Ashley P. Fidler:** Conceptualization (supporting); Data Curation (lead); Methodology (equal); Investigation (lead); Formal analysis (supporting); Software (lead); Validation (equal); Visualization (lead); Writing/Original Draft Preparation (lead); Writing/Review & Editing (equal)

**Liying Chen:** Data Curation (supporting); Methodology (supporting); Investigation (supporting); Formal analysis (lead); Software (supporting); Validation (equal); Writing/Review & Editing (supporting)

**Alexander M. McKillop:** Investigation (supporting); Formal analysis (supporting); Software (supporting); Writing/Review & Editing (supporting)

**Marissa L. Weichman:** Conceptualization (lead); Funding acquisition (lead); Methodology (equal); Project administration (lead); Resources (lead); Supervision (lead); Writing/Review & Editing (equal)

**Data availability**

The data that support the findings of this study are available from the corresponding author upon reasonable request.

# - SUPPLEMENTARY MATERIAL -

# Ultrafast dynamics of CN radical reactions with chloroform solvent under vibrational strong coupling

Ashley P. Fidler,[1] Liying Chen,[1] Alexander M. McKillop,[1] and Marissa L. Weichman[1,a)]

[1]Department of Chemistry, Princeton University, Princeton, New Jersey 08544, USA

[a)]weichman@princeton.edu

**Supplementary Section S1: Validation of transient absorption data processing procedures in extracavity measurements**

      Before performing intracavity experiments in the CN + CHCl$_3$ reaction system, we obtain extracavity transient absorption (TA) measurements in 56 μm pathlength cells fitted with infrared (IR)-transparent calcium fluoride (CaF$_2$) windows. We use these measurements to verify that the time constants we extract for free CN radical population decay and CN-solvent complexation are consistent with literature values. We analyze all TA data according to procedures described in the ICN photodissociation literature.[1,2] Briefly, we record the white light (WL) probe spectrum as a function of pump-probe time delay after UV pump-induced photolysis of ICN. Broad spectral features are observed at 390 nm and 340 nm. To extract time constants for these features, we perform a multiple variable regression with three components: (i) a Lorentzian profile fit at early pump-probe delays for the free CN radical population that absorbs at 390 nm, (ii) a late pump-probe delay spectrum for the blue-shifted CN-solvent complexes that appear at 340 nm, and (iii) a broad "solvent" feature composed from the post-overlap neat CHCl$_3$ TA spectrum recorded in an identical CaF$_2$ cell that represents not only absorption from a residual CHCl$_3$ photoproduct, but also a I($^2$P$_{3/2}$)-solvent charge transfer band. The amplitudes of the 390 nm and 340 nm regression components are plotted as function of delay and fit to a biexponential decay or an exponential growth followed by a decay, respectively. See Section III.B of the main text for more details on the three-component regression and dynamic processes underlying the evolution of the spectral features. As shown in Table I of the main text, our measured extracavity time constants are consistent with those reported in the literature.

      Spectral chirp originating from WL generation is apparent in our broadband TA measurements. Fig. S1(a) shows representative extracavity TA data for a 0.5 M ICN solution in

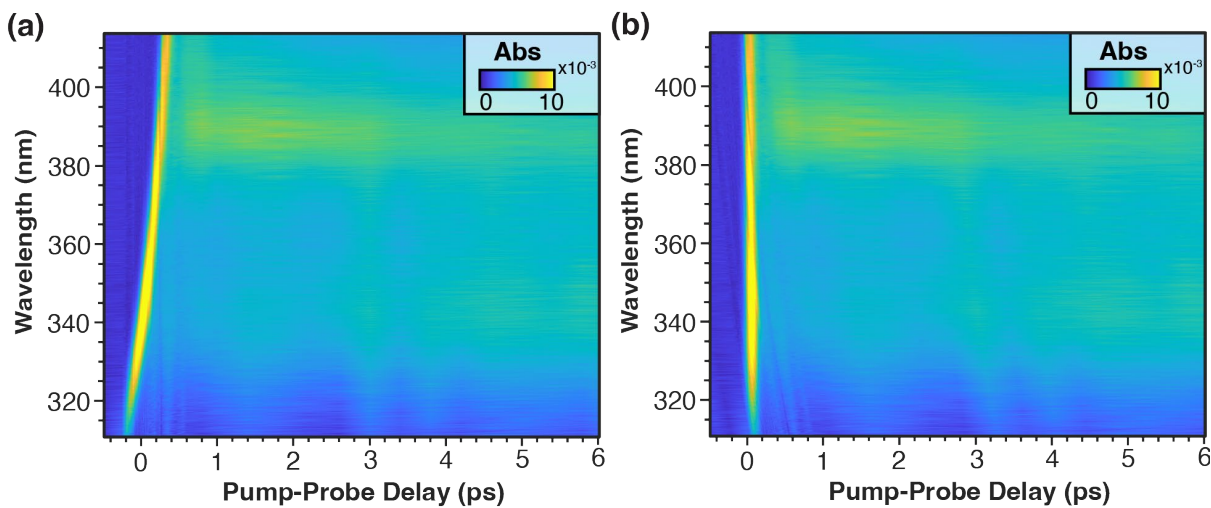

**Fig. S1.** Transient absorption data for a 0.5 M ICN solution in CHCl$_3$ as measured through a 56 μm pathlength CaF$_2$ cell **(a)** before and **(b)** after chirp correction. The TA spectra are plotted as a function of wavelength and pump-probe delay. Warmer colors correspond to two-photon absorption signal at temporal overlap and absorption signals due to the emergence of transient features associated with the formation of new species following photolysis.

CHCl$_3$ near pump-probe overlap. The delay at which the pump and probe beams are temporally overlapped varies with wavelength, as demonstrated by the intense two-photon absorption feature at time zero.[3] We implement the post-processing protocols of Grubb et al.[1] to correct for this spectral chirp. We fit a Gaussian function to the temporal profile of the two-photon absorption feature at each wavelength and use the maximum of the first derivative to determine time overlap ($t_0$). Time overlap is then plotted as a function of wavelength and fit to the expression for spectral dispersion:[1]

$$t_0(\lambda) = ABP + GDD \left(\frac{2\pi c}{\lambda}\right) + \frac{1}{2} TOD \left(\frac{2\pi c}{\lambda}\right)^2 \quad (S1)$$

where $ABP$ is the absolute spectral phase, $GDD$ is the group delay dispersion, $TOD$ is the third order dispersion of the pulse, and $c$ is the speed of light. We use this fit to recalibrate the pump-probe time delay axis at each wavelength to account for spectral chirp, as shown in Fig. S1(b). Note that we must interpolate the experimental data in the time-domain to ensure consistent time delay points among all spectra. After applying the chirp correction, we implement the three-component regression and fit the resulting time traces as described above.

We now compare the extracavity time constants for the 390 nm and 340 nm features obtained with and without chirp correction. Temporal data and fits without chirp correction are shown in Fig. S2(a,b), and described further in Section III.B of the main text. In this case, the extracavity time profiles for the 390 nm and 340 nm features are well-fit by a biexponential decay, and by a fast exponential growth followed by a slow exponential decay, respectively. The temporal profiles obtained *using* chirp correction for the 390 nm and 340 nm spectral features (Fig. S2(c,d)) include more interpolated data points but otherwise do not noticeably differ from their uncorrected counterparts. As shown in Table SI, the time constants extracted for the chirp-corrected regression components are statistically equivalent to those obtained without chirp correction. We therefore refrain from implementing any chirp correction procedures on the TA data presented here, especially given the propensity of chirp correction to introduce artifacts from experimental noise between time delay points into the spectral traces, and therefore into our regression.[1]

**Table SI:** Comparison of CN + CHCl$_3$ time constants obtained in a 56 µm pathlength CaF$_2$ cell with and without chirp correction procedures. The error bars represent uncertainty in the fit.

|  | 390 nm | | 340 nm | |
| --- | --- | --- | --- | --- |
| Dataset | $\tau_1$ (ps) | $\tau_2$ (ps) | $\tau_1$ (ps) | $\tau_2$ (ps) |
| Single Dataset – No Chirp Correction | 3.6 ± 0.3 | 59 ± 5 | 2.3 ± 0.3 | 1620 ± 120 |
| Single Dataset – Chirp Corrected | 3.7 ± 0.1 | 60 ± 1 | 2.0 ± 0.1 | 1730 ± 4 |

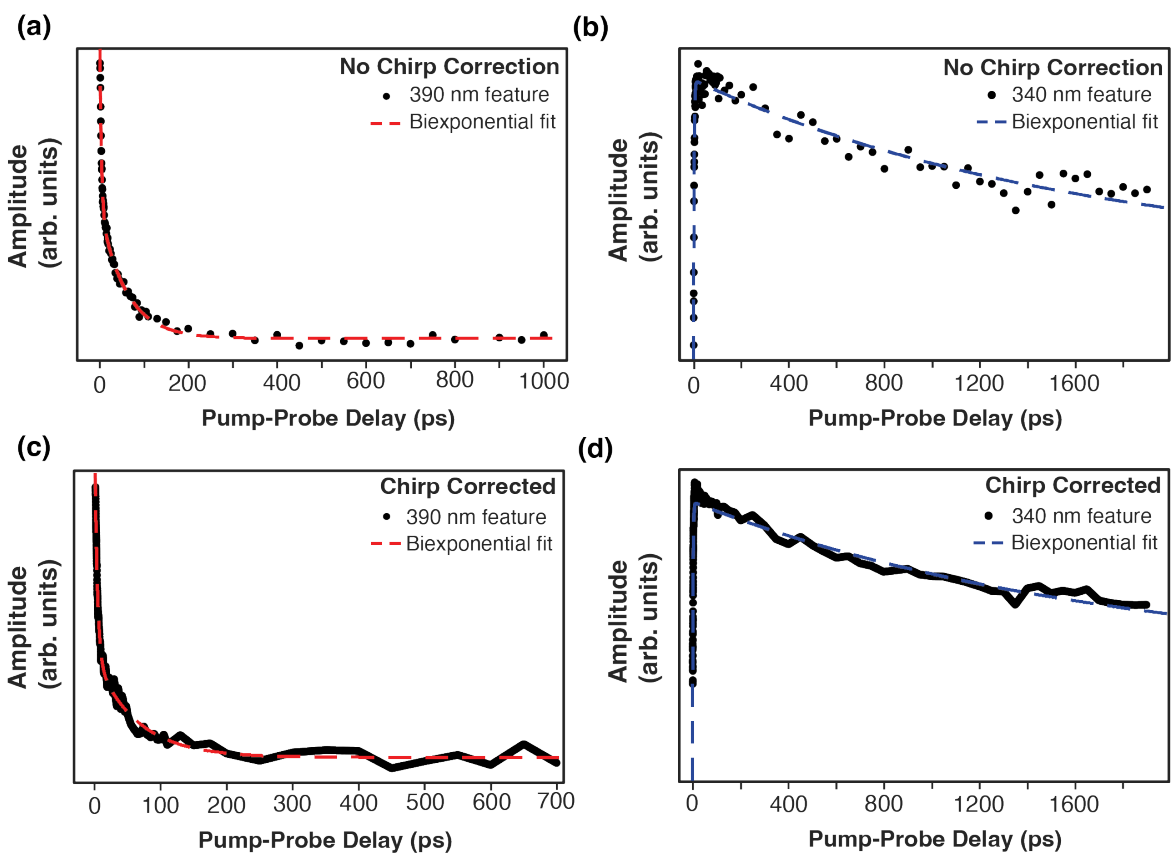

**Fig. S2.** CN + CHCl$_3$ reaction dynamics measured with and without chirp correction. Time traces and biexponential fits of the **(a)** 390 nm and **(b)** 340 nm spectral components extracted from extracavity transient measurements without any chirp correction procedures applied. Similar traces and fits are obtained for the **(c)** 390 nm and **(d)** 340 nm features after chirp correction.

**Supplementary Section S2: Cavity length dependence of infrared transmission and white light transient absorption spectra**

While polariton chemistry experiments documented in the literature typically employ thin (sub-10 µm) cavities and low mode-order longitudinal cavity modes for VSC,[4-6] we perform measurements in thicker cavities in order to maximize our TA signal amplitude. The IR transmission spectra for 25 µm and 56 µm cavities shown in Fig. S3(a) illustrate that the free spectral range (FSR) in a 56 µm cavity is sufficient to allow for clear identification of the relatively conservative Rabi splitting (25 cm$^{-1}$) obtained for the C−H stretching mode of chloroform. We do not observe unwanted spurious couplings with other vibrational resonances despite the denser cavity mode spacing because the chloroform spectrum is sparse within the narrowband high-reflectivity region of the DBR mirrors.

As reported elsewhere,[7, 8] increasing the cavity thickness does not substantially modify the collective cavity coupling strength. Therefore, we chose the cavity length on the basis of TA signal amplitude. Fig. S3(b) shows the absorbance measured at 390 nm at early pump-probe time delays for a 0.5 M ICN solution in CHCl$_3$ contained within 25 μm and 56 μm DBR cavities. Larger TA signals are clearly obtained from the thicker 56 μm cavity, although the difference in signal amplitude between the 25 μm and 56 μm cavities is not linear with cavity length due to the noncollinear beam geometry and TA artifacts from the DBR mirrors (see Supplementary Section S5). The same trend holds for the 340 nm feature (Fig. S3(c)). We therefore perform intracavity TA measurements in 56 μm cavities.

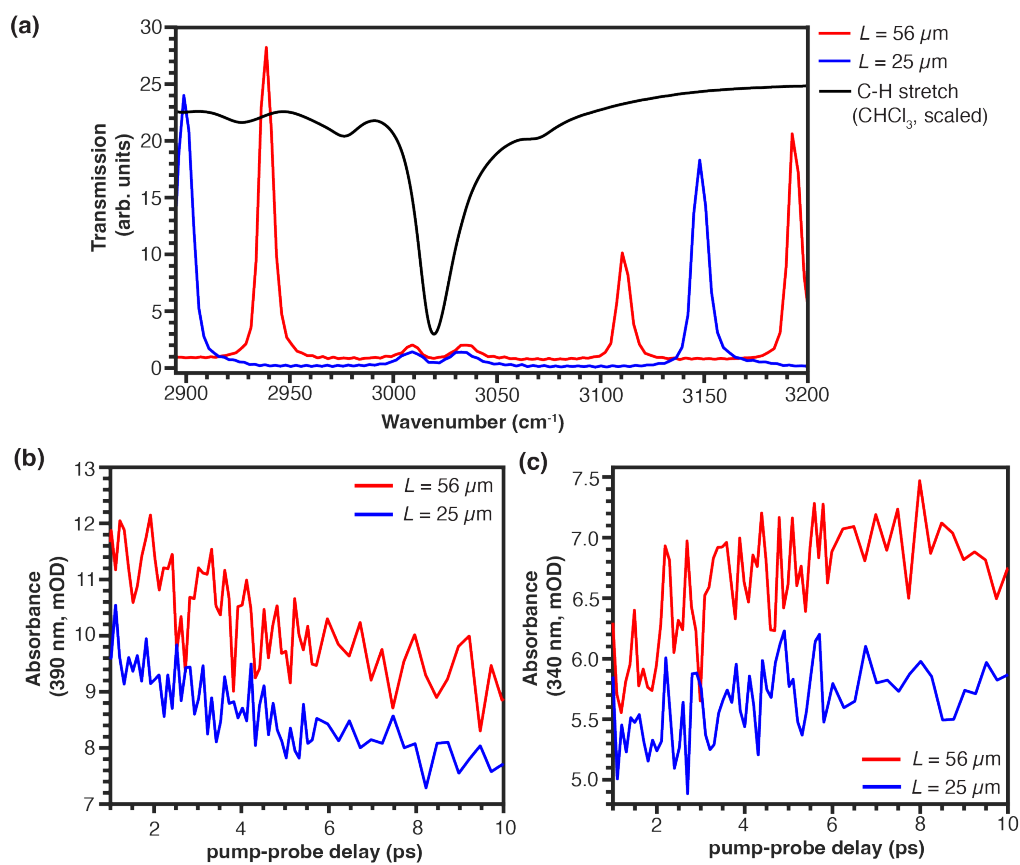

**Fig. S3.** Comparison of cavity transmission spectra and transient absorption spectra obtained in 25 μm (blue) and 56 μm (red) length cavities. **(a)** Infrared cavity transmission spectra obtained demonstrating vibrational strong coupling of the C−H stretch of CHCl$_3$. An extracavity transmission spectrum of CHCl$_3$ (black) is provided as a reference. Transient absorption traces for a 0.5 M ICN solution in CHCl$_3$ taken at **(b)** 390 nm and **(c)** 340 nm as a function of pump-probe delay.

**Supplementary Section S3: DBR mirror reflectivity**

The distributed Bragg reflector (DBR) mirrors (UltraFast Innovations GmbH) used in our experiments are fabricated with alternating layers of high and low refractive index materials ($HfO_2$, $SiO_2$) to permit optical access for our ultraviolet (UV) and WL pulses while ensuring high reflectivity in the C−H stretching region. We measure the DBR mirror reflectivity profile with an infrared imaging microscope (Nicolet iN10 Mx, Thermo Scientific). As specified by the manufacturer, the mirrors exhibit a narrow band of high reflectivity (R ≥ 90%) in the C−H stretching region between 2950 – 3365 $cm^{-1}$ (Fig. S4). At 3020 $cm^{-1}$, the frequency of the $CHCl_3$ C−H stretching mode we target for VSC, the measured mirror reflectivity is $R\sim93\%$. The reflectivity of the mirrors does not change appreciably with repeated use. We utilize this reflectivity curve to simulate the transmission spectra of Fabry-Pérot cavities composed of two of these DBR mirrors, as described in Section III.A of the main text.

**Supplementary Section S4: Alternative near-resonance cavity-coupling condition for vibrational strong coupling of the chloroform C−H stretching mode**

In the main text, we define the "on-resonance" cavity coupling condition with the $CHCl_3$ C−H symmetric stretch as the case where the two polariton peaks are of equal amplitude. Experimental and simulated examples of this on-resonance cavity coupling condition are provided in Fig. 5(c,f) of the main text. However, if we use the experimental dispersion curve provided in Fig. 5(b) to identify the cavity tuning condition with the smallest Rabi splitting, the "equal amplitude" on-resonance coupling condition does not in fact correspond to the minimum Rabi splitting. A cavity coupling condition with a less intense upper polariton peak results in a Rabi splitting of 24 $cm^{-1}$, as compared to 25 $cm^{-1}$ splitting observed in the on-resonance spectrum. This near-resonance cavity coupling position is shown in Fig. S5(a). The asymmetric appearance of this coupling condition stems from the asymmetric CHCl$_3$ vibrational lineshape, which features increased absorption on its higher frequency shoulder (see Fig. 5(a,d) of the main text).

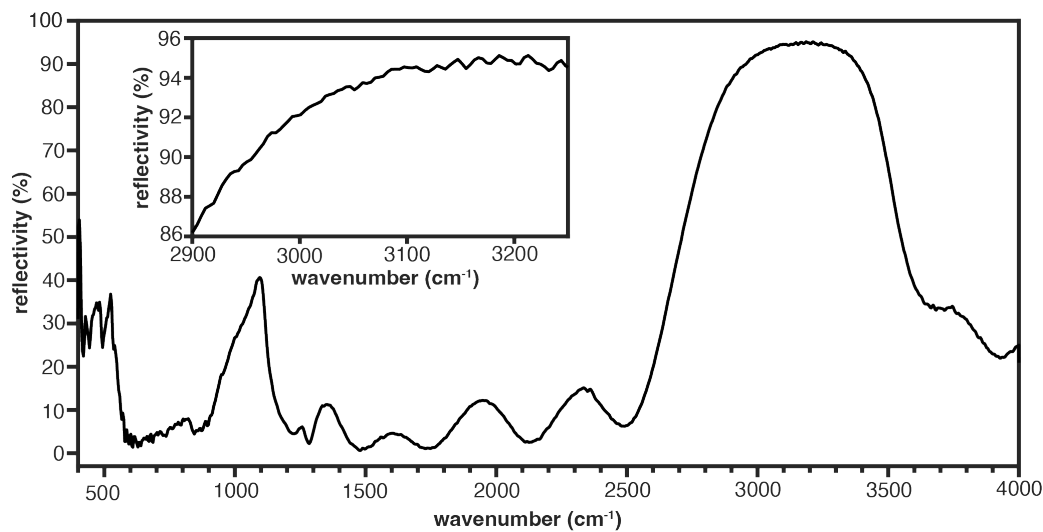

**Fig. S4.** Reflectivity spectrum measured for a single DBR mirror using an infrared imaging microscope. The inset provides the mirror reflectivity in greater detail for the C−H stretching region of interest.

Simulations confirm the experimentally identified minimum Rabi splitting cavity coupling condition. Fig. S5(b) depicts the simulated cavity transmission spectrum with the minimum Rabi splitting calculated using the classical expression for the transmission of light through a Fabry-Pérot cavity given in Eqn. 2 of the main text. The simulated Rabi splitting for this near-resonance coupling condition is 25 cm$^{-1}$, which is slightly smaller than the 26 cm$^{-1}$ splitting found for the on-resonance cavity coupling condition. The simulated and experimental spectra therefore agree that the cavity coupling condition with the minimum Rabi splitting is characterized by a less intense upper polariton peak. We refer to this coupling condition as the "near-resonance" case moving forward.

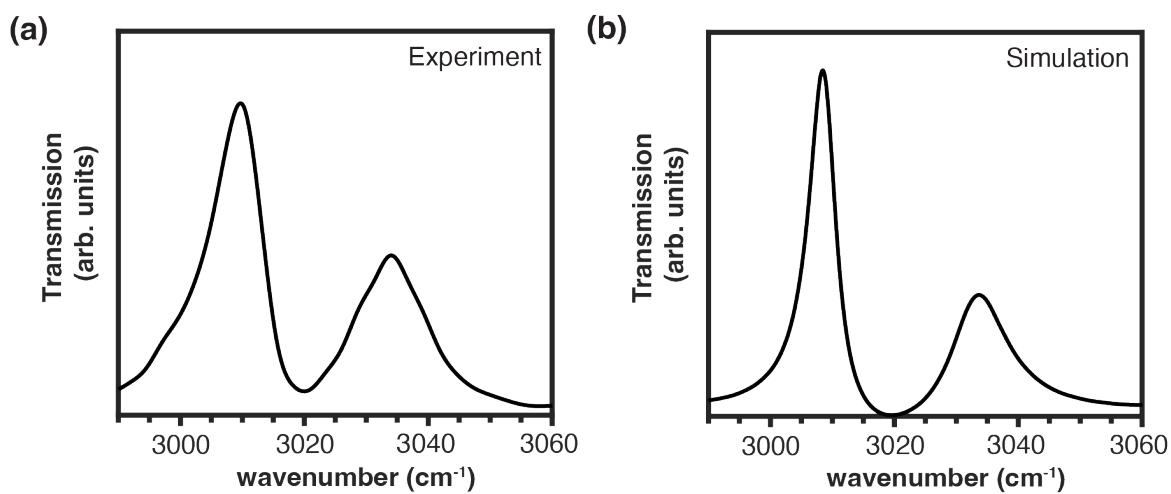

**Fig. S5.** Near-resonance condition that yields the minimum Rabi splitting for vibrational strong coupling of the CHCl$_3$ C−H stretching mode in a DBR cavity with a nominal $L$=56 μm spacer. **(a)** Experimental and **(b)** simulated cavity transmission spectra tuned to this cavity coupling condition exhibit reduced intensity of the upper polariton peak.

**Supplementary Section S5: Transient absorption artifacts in DBR mirrors**

Our DBR cavity mirrors are composed of alternating dielectric layers of high and low refractive index materials ($HfO_2$, $SiO_2$) on $SiO_2$ substrates for high reflectivity in the C−H stretching region of the IR spectrum as well as adequate transmission at UV and visible wavelengths. However, the DBR mirror coatings absorb a significant percentage of UV light, as shown in Fig. 4(b) of the main text. Transmission UV pump – white light probe TA measurements performed on a single, clean DBR mirror demonstrate broadband transient features that decay on picosecond timescales comparable to those of the CN + $CHCl_3$ reaction dynamics of interest (Fig. S6). TA measurements performed in DBR cavities will therefore inevitably incorporate transient features from both the intracavity sample and the dielectric mirror coatings. We account for these transient DBR artifacts in the multiple variable regression via a spectrally broad "solvent" component represented by a spectrum of neat $CHCl_3$ at 100 ps, as detailed in Section III.B of the main text.

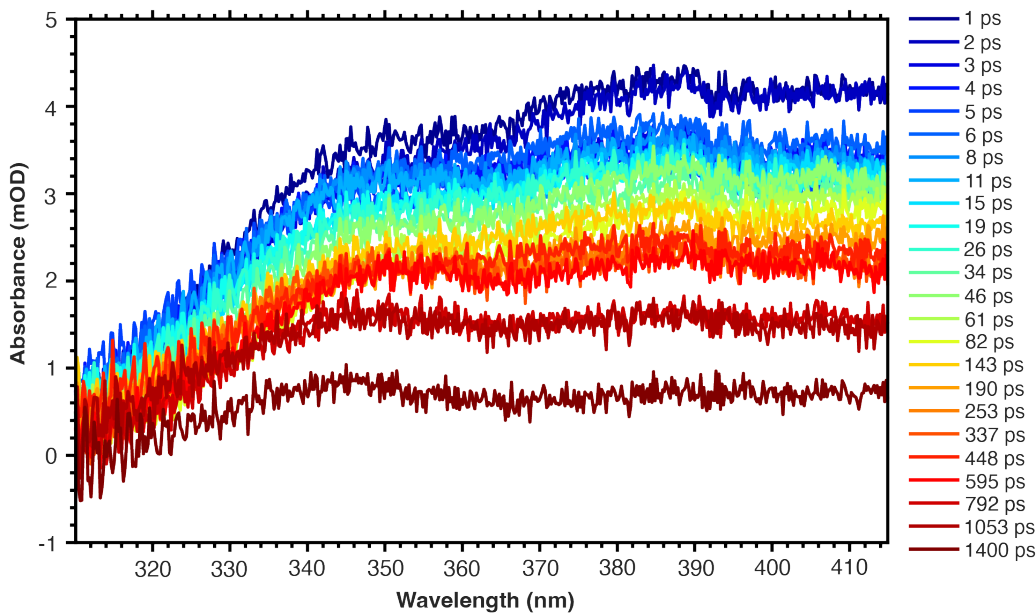

**Fig. S6.** Representative ultraviolet pump - white light probe transient absorption spectra measured in a single DBR mirror reveal spectrally broad transient features.

**Supplementary Section 6: Reproducibility of intracavity transient absorption measurements performed on- and off-resonance with the CHCl$_3$ C−H stretching mode**

To ensure the accuracy of the time constants we record for intracavity CN + CHCl$_3$ reaction dynamics, we collected multiple TA datasets in cavities tuned on- and off-resonance with the C−H stretching mode of chloroform over a period of nearly four months. We measure on- and off-resonance datasets in pairs without changing the laser alignment conditions or data collection parameters to establish a contemporaneous control for each experiment. Time constants are extracted using the procedures described in Section III.B of the main text. We eliminate any dataset pairs that meet the following criteria: (i) the regression statistics return a p-value $\geq 0.05$, indicating that the measured spectra are not accurately modelled by the three components we input into the regression, and (ii) any extracted time constants fall more than three standard deviations outside the norm, as calculated from all datasets collected at a particular cavity coupling condition. We are left with 16 pairs of on- and off-resonance transient absorption data. The time constants extracted for each individual on-resonance dataset are provided in Table SII. The corresponding off-resonance time constants measured in tandem are given in Table SIII. As shown in Table I of the main text, the average on- and off-resonance time constants are statistically equivalent.

**Table SII:** Time constants obtained for the CN + CHCl$_3$ reaction in cavities tuned to the on-resonance condition with polariton peaks of equal amplitude as shown in Fig. 5(c,f) of the main text.

| Dataset | 390 nm | | 340 nm | |
| --- | --- | --- | --- | --- |
| | $\tau_1$ (ps) | $\tau_2$ (ps) | $\tau_1$ (ps) | $\tau_2$ (ps) |
| 2022-09-13 #1 | 3.6 | 80 | 2.2 | 1900 |
| 2022-12-08 #1 | 3.8 | 101 | 2.4 | 1900 |
| 2022-12-13 #1 | 3.9 | 77 | 1.7 | 1200 |
| 2022-12-14 #1 | 3.6 | 73 | 2.0 | 1300 |
| 2022-12-14 #2 | 3.5 | 69 | 2.0 | 2200 |
| 2022-12-15 #1 | 3.7 | 79 | 2.2 | 1100 |
| 2022-12-15 #2 | 3.7 | 67 | 2.7 | 1800 |
| 2022-12-15 #3 | 3.6 | 74 | 2.1 | 1500 |
| 2022-12-20 #1 | 2.7 | 57 | 1.9 | 1400 |
| 2022-12-20 #2 | 4.1 | 79 | 1.7 | 1700 |
| 2022-12-21 #1 | 3.3 | 55 | 2.2 | 1100 |
| 2023-01-09 #1 | 2.9 | 72 | 2.4 | 1500 |
| 2023-01-09 #2 | 3.4 | 67 | 2.3 | 800 |
| 2023-01-09 #3 | 2.9 | 42 | 2.1 | 1300 |
| 2023-01-10 #1 | 2.8 | 48 | 3.3 | 2500 |
| 2023-01-11 #1 | 2.6 | 87 | 3.9 | 1700 |
| Mean | 3.4 ± 0.5 | 70 ± 15 | 2.3 ± 0.6 | 1500 ± 400 |

**Table SIII:** Time constants obtained for the CN + CHCl₃ reaction in cavities tuned off-resonance with the C−H stretch of chloroform taken in tandem with measurements performed at the on-resonant cavity coupling condition shown in Fig. 5(c,f) of the main text.

|  | *390 nm* | | *340 nm* | |
| --- | --- | --- | --- | --- |
| *Dataset* | $\tau_1$ *(ps)* | $\tau_2$ *(ps)* | $\tau_1$ *(ps)* | $\tau_2$ *(ps)* |
| 2022-09-13 #1 | 3.1 | 64 | 2.9 | 1100 |
| 2022-12-08 #1 | 3.7 | 92 | 2.5 | 1900 |
| 2022-12-13 #1 | 3.2 | 74 | 1.4 | 1300 |
| 2022-12-14 #1 | 3.6 | 77 | 2.4 | 1400 |
| 2022-12-14 #2 | 3.7 | 64 | 1.9 | 2200 |
| 2022-12-15 #1 | 3.6 | 73 | 2.7 | 1500 |
| 2022-12-15 #2 | 3.9 | 67 | 3.4 | 1900 |
| 2022-12-15 #3 | 3.8 | 79 | 1.9 | 1500 |
| 2022-12-20 #1 | 2.3 | 68 | 1.8 | 1400 |
| 2022-12-20 #2 | 2.8 | 55 | 2.1 | 1600 |
| 2022-12-21 #1 | 2.8 | 47 | 1.8 | 1600 |
| 2023-01-09 #1 | 3.3 | 88 | 2.3 | 1500 |
| 2023-01-09 #2 | 2.6 | 85 | 2.3 | 1000 |
| 2023-01-09 #3 | 2.2 | 39 | 1.8 | 1700 |
| 2023-01-10 #1 | 3.6 | 59 | 3.1 | 2400 |
| 2023-01-11 #1 | 2.7 | 100 | 3.3 | 2200 |
| Mean | 3.2 ± 0.6 | 71 ± 16 | 2.4 ± 0.6 | 1600 ± 400 |

**Supplementary Section S7: Transient absorption measurements of the CN + CHCl₃ reaction system in cavities tuned to the near-resonance cavity coupling condition**

We perform additional TA measurements on our CN + CHCl₃ reaction system at the alternative near-resonance cavity coupling condition described in Supplementary Section S4. As in the on-resonance experiments described in Supplementary Section S6, we measure near-resonance and off-resonance datasets in tandem to establish a control for each experiment. Using our three-component multiple variable regression (see Section III.B of the main text) on the near-resonance TA data, we obtain time traces for free CN radical population decay and CN-solvent complexation at 390 nm and 340 nm, respectively. Fig. S7(a) demonstrates that the 390 nm feature remains well-described by a biexponential decay due to loss of the free CN population to recombination, complexation with solvent, and hydrogen-abstraction (see Fig. 2(b) in the main text for reaction schematic). Similarly, the time trace for the 340 nm feature can still be fit by fast, exponential growth followed by a slow decay (Fig. S7(b)).

Table SIV reports time constants extracted from 16 TA datasets collected with cavities tuned to the minimum Rabi splitting near-resonance coupling condition. The average time constants are consistent with the on- and off-resonance intracavity constants reported in the main text and Supplementary Section S6 within experimental uncertainty. In addition, these near-resonance time constants are statistically identical to the time constants from nine off-resonance control measurements taken concurrently with the near-resonance data, as shown in Table SV. The relatively large error bars for the $\tau_2$ decay constant of the 340 nm feature originate from drift in spatial overlap of the pump and probe beams at the extrema of the delay line.

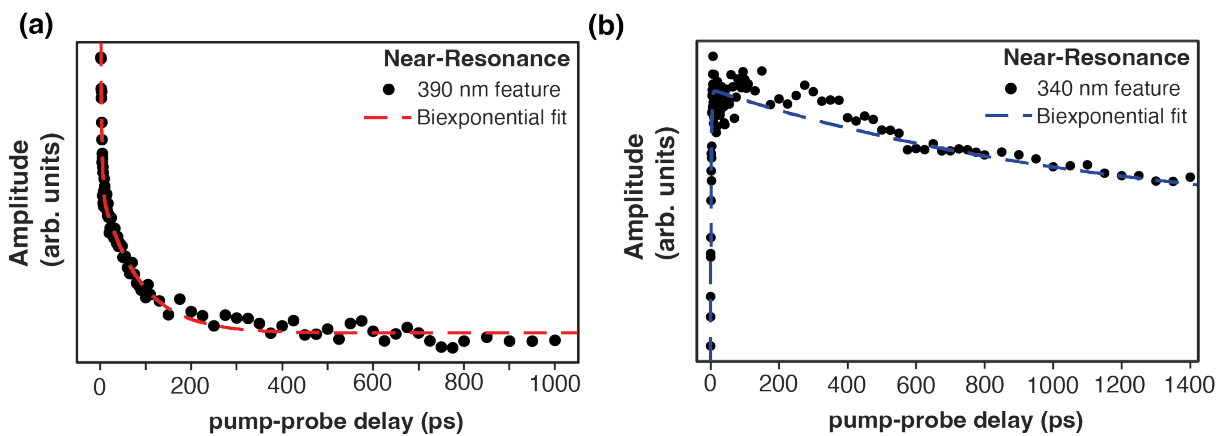

**Fig. S7.** Intracavity CN + CHCl₃ reaction dynamics measured at the near-resonance cavity coupling condition that yields the minimum Rabi splitting of the CHCl₃ C−H stretch. Representative time traces and biexponential fits of the **(a)** 390 nm and **(b)** 340 nm spectral components extracted from TA measurements performed in the 56 μm DBR cavity for a 0.5 M ICN solution in CHCl₃.

**Table SIV:** Time constants obtained for the CN + CHCl$_3$ reaction in cavities tuned to the minimum Rabi splitting near-resonant cavity coupling condition.

|  | 390 nm | | 340 nm | |
| --- | --- | --- | --- | --- |
| *Dataset* | $\tau_1$ *(ps)* | $\tau_2$ *(ps)* | $\tau_1$ *(ps)* | $\tau_2$ *(ps)* |
| 2023-05-16 #1 | 3.1 | 75 | 2.0 | 1100 |
| 2023-05-16 #2 | 2.4 | 85 | 1.8 | 1200 |
| 2023-05-16 #4 | 3.2 | 77 | 2.5 | 800 |
| 2023-05-18 #1 | 3.7 | 54 | 3.2 | 900 |
| 2023-05-18 #3 | 3.0 | 74 | 2.6 | 900 |
| 2023-05-18 #4 | 3.2 | 57 | 2.6 | 900 |
| 2023-05-18 #5 | 3.7 | 75 | 2.0 | 1100 |
| 2023-05-18 #6 | 3.3 | 83 | 2.2 | 1100 |
| 2023-05-18 #7 | 4.1 | 67 | 1.7 | 1300 |
| 2023-05-19 #1 | 3.9 | 91 | 2.0 | 800 |
| 2023-05-19 #2 | 3.9 | 101 | 2.4 | 800 |
| 2023-05-19 #3 | 4.7 | 92 | 2.7 | 1000 |
| 2023-05-25 #1 | 3.2 | 60 | 3.0 | 1300 |
| 2023-05-25 #3 | 2.5 | 64 | 3.1 | 900 |
| 2023-05-25 #4 | 2.9 | 57 | 2.8 | 1000 |
| 2023-05-25 #5 | 3.1 | 70 | 2.5 | 900 |
| Mean | 3.4 ± 0.6 | 74 ± 12 | 2.4 ± 0.5 | 1000 ± 200 |

**Table SV:** Time constants obtained for the CN + CHCl$_3$ reaction in cavities tuned off-resonance with the C-H stretch of chloroform taken in tandem with measurements performed at the minimum Rabi splitting near-resonant cavity coupling condition

|  | 390 nm | | 340 nm | |
| --- | --- | --- | --- | --- |
| *Dataset* | $\tau_1$ *(ps)* | $\tau_2$ *(ps)* | $\tau_1$ *(ps)* | $\tau_2$ *(ps)* |
| 2023-05-16 #1 | 3.7 | 94 | 2.2 | 1600 |
| 2023-05-16 #2 | 3.4 | 97 | 2.0 | 1100 |
| 2023-05-16 #4 | 3.3 | 74 | 2.1 | 1000 |
| 2023-05-18 #1 | 3.6 | 76 | 3.4 | 1700 |
| 2023-05-18 #2 | 3.2 | 57 | 2.8 | 900 |
| 2023-05-25 #1 | 3.5 | 79 | 2.9 | 1300 |
| 2023-05-25 #2 | 2.1 | 74 | 2.5 | 900 |
| 2023-05-25 #3 | 2.9 | 77 | 2.9 | 900 |
| 2023-05-25 #4 | 4.4 | 69 | 2.6 | 900 |
| Mean | 3.3 ± 0.6 | 77 ± 12 | 2.6 ± 0.5 | 1100 ± 300 |

**Supplementary Section S8: Key experimental parameters compared to literature values**

In Table SVI, we compare our reaction system and DBR cavities to those employed in prior literature experiments by Ebbesen and coworkers and Simpkins and coworkers which report VSC-altered chemistry.[6, 9] We have calculated values for the reduced coupling strength, cavity mode order, Q factor, and photon lifetime for these literature experiments where they were not provided explicitly. We compute the reduced coupling strength, $\eta$, from the Rabi frequency ($\Omega_R$) and the coupled vibrational mode frequency ($\nu_0$) according to the ratio

$$\eta = \frac{\Omega_R}{\nu_0} \quad (S1)$$

with both $\Omega_R$ and $\nu_0$ given in wavenumbers (cm$^{-1}$). We estimate the cavity mode order, $m$, using the cavity length ($L$, in µm) and the real component of the refractive index ($n$):

$$m = \frac{2nL}{10^4}\nu_0 \quad (S2)$$

The cavity Q factor is calculated from the ratio between $\nu_0$ and the cavity mode linewidth ($\Delta\nu_{FWHM}$) in wavenumbers (cm$^{-1}$):

$$Q = \frac{\nu_0}{\Delta\nu_{FWHM}} \quad (S3)$$

Using the Q factor, we can calculate the lifetime of a photon in the cavity ($\tau_p$):

$$\tau_p = \frac{Q}{2\pi c \nu_0} \quad (S4)$$

where $c$ is the speed of light.

**Table SVI:** Key experimental parameters for this work and two literature experiments that reported VSC-modified chemical reactivity.

| | This work | Ahn *et al.* (2023)[6] | Thomas *et al.* (2019)[9] |
|---|---|---|---|
| Reaction | hydrogen abstraction | phenyl isocyanate alcoholysis | silyl bond cleavage |
| Activation energy (kJ/mol) | barrierless | 28 | 24 |
| Time range monitored | 1900 ps | 7 hours | 120 mins |
| Vibrational mode | C – H stretch | NCO stretch | Si – C stretch |
| Coupled vibrational mode frequency ($\nu_0$, cm$^{-1}$) | 3020 | 2260 | 842 |
| Rabi splitting ($\Omega_R$, cm$^{-1}$) | 25 | 112 | 70 |
| Reduced coupling strength ($\eta$) | 0.008 | 0.05 | 0.08 |
| Nominal cavity length ($L$, µm) | 56 | 6 – 12 | 6 – 8 |
| Cavity mode order, $m$ | 30 – 40 | 4 – 8 | 2 |
| Cavity Q factor | 530 | 100 | 30 |
| Photon lifetime ($\tau_p$, ps) | 0.9 | 0.23 | 0.2 |